\documentclass[pra,preprint,12pt]{revtex4}
\usepackage{dcolumn}
\usepackage{mathrsfs}
\usepackage{amsmath}
\usepackage{graphicx}
\usepackage{bm}
\usepackage{epstopdf}
\usepackage{float}
\usepackage{hyperref}
\usepackage{graphicx,subfigure,xcolor}
\usepackage{braket}

\newcommand{\bea}{\begin{eqnarray}}
\newcommand{\eea}{\end{eqnarray}}
\newcommand{\beq}{\begin{equation}}
\newcommand{\eeq}{\end{equation}}

\def\/{\over}

\def\w{\omega}
\def\w0{\omega_0}

\def\br{{\bf r}}
\def\bp{{\bf p}}
\def\bA{{\bf A}}
\def\bE{{\bf E}}
\def\bB{{\bf B}}
\def\bk{{\bf k}}
\def\bkp{{\bf k}'}
\def\bkpp{{\bf k}''}
\def\wk{\omega_k}
\def\wkp{\omega_{k'}}
\def\wkpp{\omega_{k''}}
\def\akl{a_{{\bk} \lambda}}
\def\akld{a_{{\bk} \lambda}^{\dagger}}
\def\ekl{\hat{{\bf e}}_{{\bk} \lambda}}
\def\eklp{\hat{{\bf e}}_{{\bk '} \lambda '}}
\def\eklpp{\hat{{\bf e}}_{{\bk ''} \lambda ''}}
\def\ekr{e^{i \bk \cdot \br}}
\def\emkr{e^{-i \bk \cdot \br}}
\def\vac{\mid 0_{\bk j}\rangle}

\def\bj{{\bf j}}
\def\bp{{\bf p}}
\def\bR{{\bf R}}
\def\bP{{\bf P}}
\def\bmu{{\bm \mu}}
\def\bD{{\bf D}}
\def\vac{\{{0_{\bk \lambda}}\}}
\def\gd{\tilde{g}}
\def\bfm{{\bf f}}

\begin{document}

\title{Dispersion Interactions between Neutral Atoms and the Quantum Electrodynamical Vacuum}

\author{Roberto Passante$^{1,2}$}
\affiliation{$^{1}$ Dipartimento di Fisica e Chimica, Universit\`{a} degli Studi di Palermo, Via Archirafi 36, I-90123 Palermo, Italia \footnote{roberto.passante@unipa.it}}
\affiliation{$^{2}$ INFN, Laboratori Nazionali del Sud, I-95123 Catania, Italy}

\begin{abstract}
Dispersion interactions are long-range interactions between neutral ground-state atoms or molecules, or polarizable bodies in general, due to their common interaction with the quantum electromagnetic field. They arise from the exchange of virtual photons between the atoms, and, in the case of three or more atoms, are not additive. In this review, after having introduced the relevant coupling schemes and effective Hamiltonians, as well as properties of the vacuum fluctuations, we~outline the main properties of dispersion interactions, both in the nonretarded (van der Waals) and retarded (Casimir--Polder) regime. We then discuss their deep relation with the existence of the vacuum fluctuations of the electromagnetic field and vacuum energy. We describe some transparent physical models of two- and three-body dispersion interactions, based on dressed vacuum field energy densities and spatial field correlations, which stress their deep connection with vacuum fluctuations and vacuum energy. These models give a clear insight of the physical origin of dispersion interactions, and also provide useful computational tools for their evaluation. We show that this aspect is particularly relevant in more complicated situations, for example when macroscopic boundaries are present. We also review recent results on dispersion interactions for atoms moving with noninertial motions and the strict relation with the Unruh effect, and on resonance interactions between entangled identical atoms in uniformly accelerated motion.
\end{abstract}

\maketitle

\section{Introduction}
\label{sec:introduction}
Van der Waals and Casimir--Polder dispersion interactions are long-range interactions between two or more neutral atoms or molecules in the vacuum space, arising from their common interaction with the quantum electromagnetic field \cite{CP1948,Milonni1994,Power2001}. A related effect is the atom--surface Casimir--Polder force between a neutral atom and a conducting or dielectric surface \cite{BAM2000}. A complete description of such interactions requires a quantum description of both matter and radiation. These forces are non additive, that is, the interaction between three on more atoms is not simply the sum of pairwise interactions; non-additive terms are present, involving coordinates of all atoms \cite{AT43,AZ60,Milton13}. Nonadditive contributions are usually small for dilute systems, but they can become relevant for dense systems \cite{BMM99,BP15}. In this paper, we review some properties of two- and three-body dispersion interactions between neutral atoms, both in the nonretarded (van der Waals) and in the retarded (Casimir--Polder) regime, and recent advances in this subject. We will give a particular emphasis on physical models stressing their relation with the zero-point fluctuations of the electromagnetic radiation field and vacuum energy. We will also briefly review some recent results for atoms in an excited state and when a boundary condition such as a conducting plane boundary is present, as well as dispersion and resonance interactions for atoms in noninertial motion. All of these results show that zero-point field fluctuations, and their change consequent to the presence of macroscopic magnetodielectric bodies or to a noninertial motion, can~be probed through dispersion forces on atoms or, in general, polarizable bodies.
This review mainly deals with two- and three-body dispersion interactions between atoms, in several physical situations, and their physical origin in terms of the zero-point fluctuations of the electromagnetic field; other~relevant aspects of dispersion interactions, for example the effect of magnetodielectric bodies on these interactions in the framework of macroscopic quantum electrodynamics, have been reviewed in~\cite{BW07,SB08}.

This review is organized as follows. In Section \ref{sec:Hamiltonian}, we introduce the minimal and multipolar atom-field coupling schemes, which will be used to calculate the dispersion interactions. In Section \ref{sec:effective Ham}, we derive effective Hamiltonians that allow considerable simplification of the calculations, in particular for three- and many-body dispersion interactions, or in the presence of macroscopic boundaries. Section~\ref{sec:vacuum} is dedicated to a brief discussion of the vacuum fluctuations, and vacuum energy densities, of the quantum electromagnetic field. Section \ref{sec:vdW-CP} deals with the two-body van der Waals and Casimir--Polder dispersion interactions, while Section \ref{sec:three-body} is dedicated to three-body forces. In Section \ref{sec:vf}, two- and three-body dispersion interactions are discussed in detail, in relation to physical models of dispersion forces based on dressed vacuum field energy densities, and on bare and dressed vacuum field spatial correlations. In Section \ref{sec:boundaries}, these results will be extended to the case in which a reflecting boundary is present. Finally, in Section \ref{sec:acceleration}, dispersion and resonance interactions between atoms in noninertial motion are discussed, as well as recent proposals to exploit these interactions as an indirect signature of the Unruh effect. Section \ref{sec:summary} is devoted to our final remarks.

\section{Atom-Field Interaction Hamiltonian: Minimal and Multipolar Coupling}
\label{sec:Hamiltonian}

In nonrelativistic quantum electrodynamics, the (transverse) vector potential field operator, using Gauss units, is given by \cite{Compagno1995,Power1964}
\begin{equation}
\bA (\br ,t) = \sum_{\bk \lambda} \sqrt{\frac {2\pi \hbar c^2}{\wk V}} \ekl \left( \akl (t) \ekr + \akld (t) \emkr \right) ,
\label{vp}
\end{equation}
where we have used the Coulomb gauge, $\nabla \cdot \bA (\br ,t)=0$, the polarization unit vectors $\ekl \, (\lambda =1,2)$ are assumed real, $\wk=ck$ in the vacuum space, $V$ is the quantization volume, and $\akl$ and $\akld$ are respectively annihilation and creation operators satisfying the usual bosonic commutation rules. Due~to the Coulomb gauge condition $\nabla \cdot \bA =0$, we have $\bk \cdot \ekl =0$ for all $\bk$.
The transverse electric field and the magnetic field are given by
\bea
\bE_\perp (\br ,t) &=& -\frac 1c \dot{\bA}(\br ,t) =  i\sum_{\bk \lambda} \sqrt{\frac {2\pi \hbar \wk}V} \ekl \left( \akl (t) \ekr - \akld (t) \emkr \right) ,
\label{ef}  \\
\bB (\br ,t) &=& \nabla \times \bA (\br ,t) = i\sum_{\bk \lambda} \sqrt{\frac {2\pi \hbar c^2}{\wk V}} \, \bk \times \ekl \left( \akl (t) \ekr - \akld (t) \emkr \right) ,
\label{mf}
\eea
where the subscript $\perp$ indicates the transverse part.
The vacuum state is defined by $\akl \mid \vac \rangle =0$, for all $(\bk \lambda )$. In the presence of boundaries or macroscopic bodies, appropriate boundary conditions must be set on the field operators, and the dispersion relation can change too (in a photonic crystal, for~example). The vacuum state is thus strongly dependent on the presence of microscopic or macroscopic matter.

The Hamiltonian of the free field is
\begin{equation}
\label{freefieldH}
H_F =\frac 1{8\pi} \int_V d^3r \! \left[ \bE_\perp^2 (\br ) + \bB^2 (\br ) \right] = \sum_{\bk \lambda}\hbar \wk \left( \akld \akl + \frac 12 \right) .
\end{equation}

In the presence of neutral atoms, the total Hamiltonian is the sum of three terms, $H_F$, $H_\text{atoms}$ and $H_I$, respectively the field, atomic and interaction Hamiltonians. In order to obtain the full Hamiltonian and discuss different matter--radiation coupling schemes, we start with the Lagrangian formalism in the Coulomb gauge.
The generalized coordinates are the coordinates $\br_\xi$ of the charged particles (the subscript $\xi$ indicates the particles present), the vector potential $\bA (\br ,t)$ and the scalar potential $\phi (\br ,t)$. In the Coulomb gauge, only the vector potential is second-quantized, while the scalar potential is described as a classical function. The Helmholtz theorem allows a unique separation, with given boundary conditions at infinity, of the electric field $\bE (\br ,t)$ in a transverse (solenoidal) part $\bE_\perp (\br ,t)$, with
$\nabla \cdot \bE_\perp (\br ,t)=0$, and a longitudinal (irrotational) part $\bE_\parallel (\br ,t)$, with $\nabla \times \bE_\parallel (\br ,t)=0$,
\begin{equation}
\bE (\br ,t) = \bE_\perp (\br ,t) + \bE_\parallel (\br ,t); \, \, \,
\bE_\parallel (\br ,t) = - \nabla \phi (\br ,t); \, \, \,
\bE_\perp (\br ,t) = - \frac 1c \dot{\bA}(\br ,t).
\label{ef-tr+lon}
\end{equation}

The magnetic field is completely transverse, due to the Maxwell's equation $\nabla \cdot \bB (\br ,t) = 0$.

For point particles with electric charge $q_\xi$, position $\br_\xi$ and mass $m_\xi$, we define the charge and current densities
\begin{equation}
\rho (\br ,t)=\sum_\xi q_\xi \delta (\br -\br_\xi ) ; \, \, \,
\bj (\br ,t) = \sum_\xi q_\xi {\dot{\br}}_\xi \delta (\br -\br_\xi ).
\label{densities}
\end{equation}

The current density can be separated in its transverse and longitudinal parts, $\bj (\br ,t) = \bj_\perp (\br ,t) + \bj_\parallel (\br ,t)$.

The Lagrangian of the coupled system is \cite{Power1964,PT78,Craig1998}
\begin{eqnarray}
\label{Lagrangian-min}
L^{\text{min}}(\br_\xi , \bA ,\phi )&=&
\int \! d^3r {\cal L}^{\text{min}}(\br ) =
\sum_\xi \frac 12 m_\xi {\dot{\br}}_\xi^2
+ \frac 1{8\pi}\int \! d^3r \left( \frac 1{c^2} {\dot{\bA}^2(\br )} - (\nabla \times \bA (\br ))^2 \right) \nonumber \\
&+& \frac 1{8\pi} \int \! d^3r \mid \nabla \phi (\br ) \mid^2
- \int \! d^3r \, \rho (\br ) \phi (\br ) +\frac 1c \int \! d^3r \, \bj_\perp (\br ) \cdot \bA (\br ) ,
\end{eqnarray}
where ${\cal L}^{\text{min}}(\br )$ indicates the Lagrangian density. The last two terms in the second line of Equation (\ref{Lagrangian-min}) give the matter-field interaction.
This Lagrangian yields the correct Maxwell-Lorentz equations of motion. It~defines the so-called minimal coupling scheme. The momenta conjugate to $\br_\xi$ and $\bA (\br )$ are
\begin{eqnarray}
\label{conjmomem}
\bp_\xi^{\text{min}} &=& \frac{\partial L^{\text{min}}}{\partial \dot{\br}_\xi} = m_\xi {\dot{\br}}_\xi +\frac {q_\xi}c \bA (\br_\xi ) ,
\nonumber \\
{\bm \Pi}^{\text{min}} (\br ) &=& \frac{\partial {\cal L}^{\text{min}}}{\partial \dot{\bA}}= \frac 1{4\pi c^2}{\dot{\bA}(\br )} = -\frac 1{4\pi c} \bE_\perp (\br ) ,
\end{eqnarray}
while the momentum conjugate to the scalar potential $\phi (\br ,t)$ vanishes.

We can now obtain the Hamiltonian of the interacting system in the Coulomb gauge
\begin{eqnarray}
\label{Hamiltonian-min}
&\ & H^{\text{min}} (\br_\xi , \bA , \bp_\xi^{\text{min}} ,{\bm \Pi}^{\text{min}}) = \sum_\xi \bp^{\text{min}}_\xi \cdot \dot{\br}_\xi+ \int \! d^3r \, {\bm \Pi }^{\text{min}} (\br )\cdot \dot{\bA}(\br ) - L^{\text{min}}
\nonumber \\
&=& \sum_\xi \frac 1{2m_\xi}\left( \bp_\xi^{\text{min}} - \frac {q_\xi}c \bA (\br_\xi ) \right)^2
+\frac 1{8\pi} \int \! d^3r \left( (4\pi c)^2 ({\bm \Pi^{\text{min}}}(\br ))^2 +(\nabla \times \bA (\br ))^2\right) \nonumber \\
&\ & +\frac 1{8\pi} \int \! d^3r \mid \! \nabla \phi (\br ) \! \mid^2 .
\end{eqnarray}

The last term includes all electrostatic (longitudinal) interactions: in the case of neutral atoms, they are the interactions between the atomic nucleus and the electrons of the same atom, eventually included in the Hamiltonians of the single atoms, as well as electrostatic (dipole--dipole, for example) interactions between different atoms, yielding nonretarded London--van der Waals forces \cite{London1930,Craig1998}. Thus,~in  the minimal coupling scheme, electromagnetic interactions between atoms are mediated by both longitudinal and transverse fields.

In the case of two neutral one-electron atoms, $A$ and $B$, respectively located at the fixed positions $\br_A$ and $\br_B$, and within dipole approximation, Equation (\ref{Hamiltonian-min}), using Equation (\ref{conjmomem}), assumes the following form  \cite{Power1964,Craig1998}:
 \begin{eqnarray}
\label{Hamiltonian-min-2}
 H^{\text{min}} &=& H_A^{\text{min}} + H_B^{\text{min}} + H_F^{\text{min}} + H_I^{min} + H_I^{dd},
\end{eqnarray}
where $H_A^{\text{min}}$ and $H_B^{\text{min}}$ are respectively the Hamiltonian of atoms $A$ and $B$, $H_F^{\text{min}}$ is the free transverse field Hamiltonian
\begin{equation}
\label{field hamiltonian}
H_F^{\text{min}} = \frac 1{8\pi}\int \! d^3r \left( \bE_\perp^2 (\br )+ \bB^2 (\br ) \right) ,
\end{equation}
and $H_I^{min}$ and $H_I^{dd}$ are, respectively, the interaction terms of the two atoms with the transverse field and the electrostatic dipole--dipole interaction between the atoms. They are given by
{\begin{equation}
\label{inter ham trasv}
H_I^{\text{min}} = \sum_{\xi = A,B} \left( \frac e{mc}\bp_\xi \cdot \bA (\br_\xi ) + \frac{e^2}{2mc^2} \bA^2({\br_\xi}) \right) ,
\end{equation}
\begin{equation}
\label{inter ham long}
H_I^{\text{dd}} = \frac 1{R^3}\left( {\bm \mu}_A \cdot {\bm \mu}_B -3 ({\bm \mu}_A \cdot \hat{\bR}) ( {\bm \mu}_B \cdot \hat{\bR}) \right) ,
\end{equation}
where $-e$ is the electron charge, ${\bm \mu}_A = -e\br_A$ and ${\bm \mu}_B = -e\br_B$ are respectively the dipole moment operators of atoms $A$ and $B$, and $\bR$ is the distance between the two atoms.}
Extension to more-electrons atoms, and to the case of three or more atoms, is straightforward.

A different and useful form of the matter--radiation interaction can be obtained by exploiting gauge invariance; it is the so-called multipolar coupling scheme, and it is more convenient than the minimal-coupling one in evaluating radiation-mediated interactions between neutral atoms \cite{PZ57,PZ59,PT78,Woolley1971,CTDG89}. The advantage of the multipolar coupling scheme is that the interaction is expressed in terms of the transverse displacement field, and not the vector potential, and that the electrostatic interaction between the atoms is embedded in the coupling with the transverse fields. This in general allows considerable simplification of the calculations and a clear physical picture.
The passage from the minimal to the multipolar coupling scheme can be obtained in two different, and equivalent, ways: by adding a total time-derivative to the Lagrangian or through a unitary transformation on the Hamiltonian \cite{PZ57,PZ59,AW70,Woolley1971,PT78,BL83,Andrews2018}. In view of the relevance to dispersion interactions, we will now briefly outline this procedure, within dipole approximation.

We first introduce the atomic polarisation operator, in electric dipole approximation, assuming that the (multi-electron) atoms are well separated, i.e., their distance is much larger that the Bohr's~radius,
\begin{equation}
\label{polarization}
\bP (\br ) = -\sum_{\xi} e_{\xi} \br_{\xi} \delta (\br -\br_{\xi} ) = \sum_\ell \bmu_\ell \delta (\br - \br_\ell ) ,
\end{equation}
where the subscript $\xi$ indicates the atomic electrons, with positions $\br_{\xi}$ with respect to their atomic nucleus. We have also introduced the atomic dipole moment operator of atom $\ell$, $\bmu_\ell = -\sum_{\xi_\ell}e_{\xi_\ell} \br_{\xi_\ell}$, with $\br_{\xi_\ell}$ the positions of the electrons of atom $\ell$. The longitudinal field of the bound charges in the atoms can be expressed as
\begin{equation}
\label{longitudinal field}
\bE_\parallel (\br ) = -\nabla \phi (\br ) = -4\pi \bP_\parallel (\br ) ,
\end{equation}
where $\bP_\parallel (\br )$ is the longitudinal part of the polarization operator, $\bP (\br ) = \bP_\perp (\br ) + \bP_\parallel (\br )$. In addition, in the electric dipole approximation,
\begin{equation}
\label{relP-j}
\bj (\br , t) = \dot {\bP}(\br ,t) .
\end{equation}

The new Lagrangian, where a total time-derivative has been added to (\ref{Lagrangian-min}), is given by
\begin{eqnarray}
\label{Lagrangian-mult}
L^{\text{mult}} &=&  \int \! d^3r {\cal L}^{\text{mult}}(\br ) =  L^{\text{min}} -\frac 1c \frac d{dt} \int \! d^3r \bP (\br ) \cdot \bA (\br ) \nonumber \\
&=& \sum_\xi \frac 12 m_\xi {\dot{\br}}_\xi^2
+ \frac 1{8\pi}\int \! d^3r \left( \frac 1{c^2} {\dot{\bA}^2(\br )} - (\nabla \times \bA (\br ))^2 \right)
+ \frac 1{8\pi} \int \! d^3r \mid \nabla \phi (\br ) \mid^2
 \nonumber \\
&\ & - \int \! d^3r \, \rho (\br ) \phi (\br )- \frac 1c \int \! d^3r  \bP_\perp (\br ) \cdot \dot{\bA} (\br ) ,
\end{eqnarray}
where ${\cal L}^{\text{mult}}(\br )$ is the new Lagrangian density, and we have used Equations (\ref{Hamiltonian-min}) and (\ref{longitudinal field}). The subscript $\xi$ identifies the (point-like) atoms located at position $\br_\ell$. The new conjugate momenta are
\begin{eqnarray}
\label{conjmome}
\bp_\xi^{\text{mult}} &=& \frac{\partial L^{\text{mult}}}{\partial \dot{\br}_\xi} = m_\xi {\dot{\br}}_\xi ,
\nonumber \\
{\bm \Pi}^{\text{mult}} (\br ) &=& \frac{\partial {\cal L}^{\text{mult}}}{\partial \dot{\bA}}= \frac 1{4\pi c^2}{\dot{\bA}(\br )} - \frac 1c \bP_\perp (\br ) = -\frac 1{4\pi c} \bD_\perp (\br ),
\end{eqnarray}
where $\bD_\perp (\br )$ is the transverse part of the electric displacement field $\bD (\br )= \bE (\br )+4\pi \bP (\br )$. Comparison of (\ref{conjmome}) with (\ref{conjmomem})  shows that the momenta conjugate to the generalized coordinates, on which canonical commutation relations are imposed in the quantization procedure, are different in the two coupling schemes. In the multipolar coupling, the momentum conjugate to the vector potential has a contribution from the atomic polarization vector.

From the Lagrangian (\ref{Lagrangian-mult}), and using (\ref{conjmome}), we obtain the Hamiltonian in the multipolar scheme within the dipole approximation,
\begin{eqnarray}
\label{Hamiltonian-mult}
H^{\text{mult}} (\br_\xi , \bA , \bp_\xi^{\text{mult}} ,{\bm \Pi}^{\text{mult}}) &=& \sum_\xi \bp_\xi^{\text{mult}} \cdot \dot{\br}_\xi+ \int \! d^3r {\bm \Pi }^{\text{mult}} (\br )\cdot \dot{\bA}(\br ) - L^{\text{mult}}
\nonumber \\
&=& \sum_\ell \left( \frac 1{2m} \sum_{\xi_\ell}(\bp_{\xi_\ell}^{\text{mult}})^2 + V_\ell \right) + \frac 1{8\pi} \int \! d^3r \left( \bD_\perp^2 (\br ) + (\nabla \times \bA (\br ))^2 \right)
\nonumber \\
&\ & -\sum_\ell \bmu_\ell \cdot \bD_\perp (\br_\ell ) +2\pi \sum_\ell \int \! d^3r \mid \bP_{\perp \ell} (\br ) \mid^2 ,
\end{eqnarray}
where $V_\ell$ contains the electrostatic nucleus--electrons interaction of each atom. The term containing the integral of $\bP_{\perp \ell}^2 (\br )$ (the squared transverse polarization vector of atom $\ell$) contributes only to single-atom self-interactions, and does not contribute to the interatomic interactions we will deal with in the next sections. Using the Hamiltonian (\ref{Hamiltonian-mult}), all interatomic interactions are mediated by the transverse field only, without electrostatic interatomic terms.

The passage from the minimal to the multipolar coupling scheme can be equivalently accomplished through a unitary transformation, the so-called Power--Zienau--Woolley transformation, given by \cite{PZ57,PZ59, Woolley1971,AW70,PT78,Craig1998, Salam10}
\begin{equation}
\label{PZW}
S = \exp \left(\frac i{\hbar c}\int \! d^3r \bP_\perp (\br ) \cdot \bA (\br ) \right) .
\end{equation}

The action of this transformation on the transverse electric field operator is
\begin{equation}
\label{Etrasf}
S^{-1} \bE_\perp (\br ) S = \bE_\perp (\br ) + 4\pi \bP_\perp (\br ) = \bD_\perp (\br ) ,
\end{equation}
where the equal-time commutator between the transverse electric field and the vector potential has been used
\begin{equation}
\label{AEcomm}
\left[ A_i (\br , t), E_{\perp j} (\br ', t) \right] = -4\pi i c \hbar \delta^\perp_{ij} (\br - \br ') ,
\end{equation}
which can be obtained from the expressions (\ref{vp}) and (\ref{ef}) of the field operators, and where $ \delta^\perp_{ij} (\br )$ is the transverse Dirac delta function. Transverse and longitudinal Dirac delta functions, which allow for extracting the transverse and longitudinal parts of a vector field, are defined as
\begin{eqnarray}
\label{traslongdelta}
\delta_{ij}^{\parallel}(\br ) &=& \frac 1{(2\pi )^3}\int \! \! d^3k \hat{k}_i \hat{k}_j \ekr ,
\nonumber \\
\delta_{ij}^{\perp}(\br ) &=& \frac 1{(2\pi )^3}\int \! \! d^3k \left( \delta_{ij} - \hat{k}_i \hat{k}_j \right) \ekr .
\end{eqnarray}

In addition,
\begin{equation}
\label{traslongdelta2}
\delta_{ij} \delta (\br ) = \delta_{ij}^{\parallel}(\br ) + \delta_{ij}^{\perp}(\br ) .
\end{equation}

When comparing the expressions of the minimal coupling Hamiltonian (\ref{Hamiltonian-min}) with the multipolar coupling Hamiltonian (\ref{Hamiltonian-mult}), some important considerations are in order. As already mentioned, in the multipolar scheme, the momentum field  conjugate to the vector potential, the transverse displacement field, is different than that in the minimal coupling scheme (i.e., the transverse electric field). When~these Hamiltonians are quantized, the photons in the two schemes are thus different objects. In addition, in the multipolar scheme, the transverse displacement field and the magnetic field appear in the multipolar field Hamiltonian, while the minimal free-field Hamiltonian is expressed in terms of the transverse electric field and the magnetic field. Finally, only the transverse displacement field appears in the interaction Hamiltonian, as shown in Equation (\ref{Hamiltonian-mult}); no electrostatic interactions are present in the interaction term, contrarily to minimal coupling case (\ref{Hamiltonian-min-2}).  When the multipolar Hamiltonian is used, all interactions among atoms are mediated by the quantized transverse fields only, and this makes the multipolar coupling scheme much more convenient for calculating dispersion and resonance interactions between atoms. Another important advantage of the multipolar scheme is that the transverse displacement field $\bD_\perp (\br )$ is fully retarded, contrarily to the transverse electric field $\bE_\perp (\br )$, that, as it is well known, contains a non-retarded part \cite{PZ59,Bykov93}. This is also evident from the fact~that
\begin{eqnarray}
\label{relazD-E}
\bD_\perp (\br ) &=& \bE_\perp (\br ) + 4\pi \bP_\perp (\br ) = \bE_\perp (\br ) + 4\pi \bP (\br ) - 4\pi \bP_\parallel (\br ) = \bE_\perp (\br ) + \bE_\parallel (\br )  + 4\pi \bP (\br )
\nonumber \\
&=&  \bE (\br )  + 4\pi \bP (\br ) ,
\end{eqnarray}
where we have used Equation (\ref{longitudinal field}). Taking into account Equation (\ref{polarization}), at points $\br$ different from the positions $\br_\ell$ of the (point-like) atoms, we have
\begin{equation}
\label{relazD-Ea}
\bD_\perp (\br ) =  \bE (\br )  \, \, \, \, \, (\br \neq \br_\ell ) .
\end{equation}

In other words, the transverse displacement field in all points of space but the atomic positions coincides with the total (transverse plus longitudinal) electric field, which is a retarded field. This is particularly relevant when issues related to field propagation and causality are considered \cite{Bykov93,BCPPP90,CPPP95}.

\section{Effective Hamiltonians}
\label{sec:effective Ham}

Dispersion interactions between two atoms (van der Waals and Casimir--Polder) are effects starting at fourth order in the radiation-matter coupling, and three-body dispersion interactions start at sixth-order \cite{Salam10,BuhmannI-2012,BuhmannII-2012,Salam08,Salam2016}. Application of fourth- or higher-order perturbation theory yields a large number of diagrams, and considerable amount of calculations.  It is therefore desirable obtaining an effective Hamiltonian that could simplify calculations, reducing the number of diagrams. In this section, we introduce an effective Hamiltonian, with an interaction term quadratic in the atom-field coupling (i.e., the electric charge), which thus allows to halve the perturbative order necessary to evaluate dispersion interactions. For instance, it allows for evaluating the two-body dispersion potential through a second-order calculation and the non-additive three-body potential through a third-order calculation, in the case of ground-state atoms or molecules. This effective Hamiltonian is expressed in terms of the dynamical polarizability of the atoms.

We start with the multipolar Hamiltonian (\ref{Hamiltonian-mult}), considering for simplicity only one atom ($A$), and~neglecting the term quadratic in the atomic transverse polarization vector, which does not play any role for the considerations that follow. Our Hamiltonian is thus
$H=H_A+H_F- \bmu_A \cdot \bD_\perp (\br_A)$, where $\br_A$ is the position of the atom. We now perform the following transformation on the Hamiltonian

\begin{eqnarray}
\label{transformation}
H_{\text{new}} &=& e^{iQ}He^{-iQ}= H_A + H_F -\bmu_A \cdot \bD_\perp (\br ) +i[ Q, H_A + H_F ] -\frac 12 [Q,[Q,H_A+H_F ]]
\nonumber \\
&+& i [Q, -\bmu_A \cdot \bD_\perp (\br_A)] + \dots ,
\end{eqnarray}
where we have used the Baker--Hausdorff relation, and the operator $Q$ is chosen in such a way to eliminate the term linear in the coupling, that is,

\begin{equation}
\label{operator}
i[Q, H_A + H_F]= \bmu_A \cdot \bD_\perp (\br_A).
\end{equation}

This condition allows for obtaining the generic matrix elements of $Q$ between atom+field states, except on the energy shell where they are undefined. Substitution of (\ref{operator}) into (\ref{transformation}) finally yields an effective Hamiltonian that is quadratic in the electric charge because the linear terms have been eliminated by the transformation. For ground-state atoms with random orientations, the effective Hamiltonian has an interaction term of the following form \cite{PPT98,PP87}
\begin{equation}
\label{EffHam}
H_{\text{int}}=-\frac 12 \sum_{\bk \lambda} \alpha_A (k) \bD_\perp (\bk \lambda , \br_A) \cdot \bD_\perp (\br_A),
\end{equation}
where $\bD_\perp (\bk \lambda , \br )$ are the Fourier components of the transverse displacement field,
$ \bD_\perp (\br )= \sum_{\bk \lambda}\bD_\perp (\bk \lambda , \br )$,
and
\begin{equation}
\label{dynpol}
\alpha_A (k) = \frac 23 \sum_m \frac {E_{mg}\mid \bmu_A^{mg} \mid^2}{E_{mg}^2-\hbar^2c^2k^2}
\end{equation}
is the dynamical polarizability of the isotropic atom. In Equation (\ref{dynpol}), $g$ indicates the ground state of the atom with energy $E_g$, $m$ a generic atomic state with energy $E_m$. In addition, $E_{mg}=E_m-E_g$, and $\bmu_A^{mg}$ are matrix elements of the dipole moment operator between states $m$ and $g$.

The effective Hamiltonian (\ref{EffHam}) has a clear classical-type interpretation: any $(\bk \lambda )$ component of the field induces a dipole moment in the atom with Fourier components determined by its dynamical polarizability $\alpha_A(k)$ at the frequency $ck$ (linear response); this induced dipole moment, then, interacts with the field at the atom's position (the factor $1/2$ in (\ref{EffHam}) appears because it is an interaction energy between induced, and not permanent, dipoles).

If all relevant field modes have a frequency $\wk =ck$ much smaller than the relevant atomic transition frequencies, that is, $ck \ll E_{mg}/\hbar$, in  (\ref{EffHam}), after taking into account the expression (\ref{dynpol}), we~can replace the dynamical polarizability with the static one, $\alpha_A = \alpha_A (0)$, obtaining the so-called Craig--Power effective Hamiltonian \cite{CP69}
\begin{equation}
\label{EffHamSt}
H_{\text{int}}=-\frac 12 \alpha_A D_\perp^2 (\br_A).
\end{equation}

The effective Hamiltonian (\ref{EffHamSt}) is valid whenever only low-frequency field modes are relevant, for example in the case of dispersion interactions between ground-state atoms in the far zone (retarded Casimir--Polder regime).
In the case of two or more atoms, the effective interaction term in the Hamiltonian is simply the sum of those relative to the single atoms \cite{PPT98}. An important point for some following considerations in this review is that Equation (\ref{EffHamSt}) shows that, taking also into account~(\ref{relazD-Ea}), the~interaction energy with a polarizable body, in the appropriate limits, involves the total electric field. This is true when the effective interaction term can be represented by its static polarizability, that~is, when all relevant field modes have frequencies lower than relevant atomic transition frequencies, allowing us to probe the electric field energy density $E^2(\br )/8\pi$, through the far-zone retarded Casimir--Polder interaction energy. A similar consideration holds for the magnetic energy density too. This will be an essential point for our following discussion in Section \ref{sec:endens}  on the relation between retarded Casimir--Polder interactions and vacuum field energies and fluctuations.

\section{Vacuum Fluctuations}
\label{sec:vacuum}

A striking consequence of the quantum theory of the electromagnetic field is the existence of zero-point field fluctuations. In the ground state of the field, where the number of photons is zero, the electric and magnetic field have quantum fluctuations around their average value zero \cite{Milonni1994}. This~means~that
\begin{equation}
\label{flut}
\langle \vac \mid \bE_\perp^2 (\br ,t) \mid \vac \rangle \neq 0 ; \,\,\,\,\, \langle 0 \mid \bB^2 (\br ,t) \mid \vac \rangle \neq 0,
\end{equation}
while $\langle \vac \mid \bE_\perp (\br ,t) \mid \vac \rangle = \langle \vac \mid \bB (\br ,t) \mid \vac \rangle = 0$, where $\mid \vac \rangle$ indicates the vacuum state of the field. Field fluctuations are directly related to the field energy density, of course. In addition, we~have
\begin{equation}
\langle \vac \mid \int \! d^3r \left[ \bE_\perp^2 (\br ) + \bB^2 (\br ) \right] \mid \vac \rangle = \sum_{\bk \lambda} \frac 12 \hbar \wk .
\label{zpe}
\end{equation}

Electric and magnetic zero-point fluctuations are a direct consequence of the non-commutativity, at equal time, of specific components of the transverse electric and magnetic field operators, given by
\begin{equation}
\label{EBcommutator}
\left[ E_{\perp m} (\br ,t), B_n(\br ' ,t) \right] = -4\pi i\hbar c \epsilon_{mn\ell} \frac \partial{\partial r_\ell} \delta (\br -\br ') ,
\end{equation}
where $\epsilon_{mn\ell}$ is the totally antisymmetric symbol, and the Einstein convention of repeated symbols has been used. Equation (\ref{EBcommutator}) can be directly obtained from the expressions (\ref{ef}) and (\ref{mf}) of the field operators.

The strength of these fluctuations, and the related field energy densities, depends on the contributions of all field modes, and thus it depends on the presence of boundary conditions, magnetodielectric bodies or matter in general.  Zero-point field fluctuations are infinite and, in the unbounded space, spatially uniform. In deriving the specific expressions of (\ref{flut}), a sum over all allowed field modes is involved and, as mentioned, the allowed modes and the dispersion relation depend on the presence of boundaries, for example metallic or magnetodielectric bodies or even single atoms or molecules, or a structured environment. Although vacuum fluctuations diverge, their difference for two different configurations of the boundaries is usually finite, and this leads to the Casimir effect, which is a tiny force between two neutral macroscopic bodies (two parallel conducting plates, in the original formulation of the effect), which has no classical analogue \cite{Casimir48,Milonni1994}. In other words, the zero-point energy is boundary-conditions-dependent, and can be varied by changing the boundary conditions. The~presence of atoms or molecules also changes the vacuum electric and magnetic energy density, and~we will see in the next sections that this is deeply related to van der Waals and Casimir--Polder forces between neutral atoms, or between atoms and macroscopic objects. This point gives clear insights on the physical origin of such forces of a pure quantum origin.

Another relevant property of zero-point field fluctuations, strictly related to Casimir--Polder interactions, is their spatial correlation. Using expression (\ref{ef}) of the transverse electric field operator, the equal-time spatial correlation function for the $(\bk \lambda)$ component of the free transverse electric field  is
\begin{equation}
\label{correl-mode}
\langle \vac \mid  E_{\perp i} (\bk \lambda ,\br ',t)  E_{\perp j} (\bkp , \lambda ',\br '',t) \mid \vac \rangle = \frac {2\pi \hbar c}V (\ekl )_i (\ekl )_j k e^{i\bk \cdot \br} \delta_{\bk \bkp} \delta_{\lambda \lambda '},
\end{equation}
where $\br = \br '-\br ''$ and the subscripts $i,j$ indicate Cartesian components. The equal-time spatial correlation of the electric field is \cite{BCPPP90}
\begin{eqnarray}
\label{ef-corr}
&\ &\langle \vac \mid  E_{\perp i} (\br ',t)  E_{\perp j} (\br '',t) \mid \vac \rangle = \frac {\hbar c}{4\pi^2}\int_0^\infty \! \! dk k^3 \int \! \! d\Omega \left( \delta_{ij} - \hat{k}_i  \hat{k}_j \right) e^{i\bk \cdot \br}
\nonumber \\
&=& -\frac {\hbar c}{4\pi^2} \left( \delta_{ij} \nabla^2 -\nabla_i \nabla_j \right) \frac ir \int_0^\infty \! \! dk k \int \! \! d\Omega e^{i\bk \cdot \br}
= -\frac {4\hbar c}\pi \left( \delta_{ij} - 2 {\hat{r}}_i {\hat{r}}_j \right) \frac 1{r^4} ,
\end{eqnarray}
where the differential operators in (\ref{ef-corr}) act on $\br$, and the sum over polarizations has been carried out by~using
\begin{equation}
\label{polarizsum}
\sum_\lambda (\ekl )_i (\ekl )_j = \delta_{ij} - \hat{k}_i \hat{k}_i ,
\end{equation}
and we have also used $ \int \! \! d\Omega e^{i\bk \cdot \br} = 4\pi \sin (kr)/(kr)$.
Equation (\ref{ef-corr}) shows that vacuum fluctuations of the electric field have an equal-time spatial correlation scaling  with the distance as $r^{-4}$; as we will discuss in more detail later on, this implies that vacuum fluctuations are able to induce and correlate dipole moments in distant atoms, and this eventually leads to their Casimir--Polder interaction energy~\cite{PT93}.

Field fluctuations and vacuum energy densities can be modified by the presence of matter---for example, a metallic or dielectric boundary (they also change in the presence of single atoms or molecules, as we will discuss in Section \ref{sec:vf}). In such cases, they depend on the position, contrarily to the unbounded-space case, where they are uniform in space.

In the case of perfectly conducting boundaries, appropriate boundary conditions must be set on the field operators at the boundaries' surface, and the electric and magnetic free field operators assume the form \cite{Milonni1994,PT82}
\begin{eqnarray}
\bE_\perp (\br ) &=& \sum_{\bk \lambda } \bE_\perp (\bk \lambda ,\br ) = i \sum_{\bk \lambda } \sqrt {\frac {2\pi \hbar \wk}V} \bfm (\bk \lambda ,\br ) \left( \akl - \akld \right) ,
\label{elfieldbound} \\
\bB (\br ) &=& \sum_{\bk \lambda } \bB (\bk \lambda ,\br ) = i \sum_{\bk \lambda } \sqrt {\frac {2\pi \hbar c^2}{\wk V}} \nabla \times \bfm (\bk \lambda ,\br ) \left( \akl - \akld \right) ,
\label{magfieldbound}
\end{eqnarray}
where $\bfm (\bk \lambda ,\br )$ are mode functions, assumed real, which take into account the boundaries present. For an infinite perfectly conducting plate located in the plane $z=0$,  renormalized electric and magnetic energy densities (that is after subtraction of the densities found in the unbounded space) are
\begin{eqnarray}
\frac 1{8\pi} \langle \vac \mid \bE^2 (z) \mid \vac \rangle_\text{ren} &=&
\frac 1{8\pi} \langle \vac \mid \bE^2 (z) \mid \vac \rangle_\text{plate} - \frac 1{8\pi} \langle \vac \mid \bE^2 (z) \mid \vac \rangle_\text{vac}
\nonumber \\
&=& \frac {3\hbar c}{32\pi^2 z^4} ,
\label{elffplate} \\
\frac 1{8\pi} \langle \vac \mid \bB^2 (z) \mid \vac \rangle_\text{ren} &=&
\frac 1{8\pi} \langle \vac \mid \bB^2 (z) \mid \vac \rangle_\text{plate} - \frac 1{8\pi} \langle \vac \mid \bB^2 (z) \mid \vac \rangle_\text{vac}
\nonumber \\
&=& -\frac {3\hbar c}{32\pi^2 z^4} ,
\label{magffplate}
\end{eqnarray}
where $z$ is the distance from the plate \cite{FS98}. Equations (\ref{elffplate}) and (\ref{magffplate}) show that the presence of the (perfectly conducting) boundary decreases the electric zero-point energy density in all space, while it increases the magnetic one. In Section \ref{sec:endens}, we will show that something similar occurs in the space around an electric dipolar source of the electromagnetic field. In the expressions above, the~renormalized field energy densities diverges at the surface of the conducting plate, $z=0$. This~is due to the unrealistic assumption of a perfectly conducting boundary with a fixed position in space; it has been shown that these divergences can be cured by assuming a fluctuating position of the plate \cite{FS98}, or introducing an appropriate cut-off to simulate a non-ideal metal boundary \cite{BP12,Bartolo2016}, thus~allowing to investigate the structure of field fluctuations in the very vicinity of the boundary \cite{BP12}. Field fluctuations near a conducting boundary that is allowed to move, with its translational degrees of freedom treated quantum mechanically and thus, with quantum fluctuations of its position, have been recently considered for both the scalar and the electromagnetic field, also in relation with the surface divergences of energy densities and field fluctuations \cite{BP13,AP15,Armata17}.

In the next sections, we will discuss how vacuum energy densities, and their space dependence in the presence of matter are strictly related to van der Waals and Casimir--Polder interactions between atoms and to the atom--surface Casimir--Polder interaction. These considerations will allow us to get a clear physical interpretation of such pure quantum interactions, as well as useful tools to evaluate them in different situations.

\section{The Van Der Waals and Casimir--Polder Dispersion Interaction between Two Neutral Ground-State Atoms}
\label{sec:vdW-CP}

Let us consider two ground-state neutral atoms, A and B, located at $\br_A$ and $\br_B$ respectively, and be $\br = \br_B - \br_A$ their distance. We assume that the atoms are in the vacuum space at zero temperature, and~at a distance such that there is no overlap between their electronic wavefunctions. We describe their interaction with the quantum electromagnetic field in the multipolar coupling scheme. \mbox{From Equation (\ref{Hamiltonian-mult}),} the Hamiltonian of the system in the multipolar coupling scheme is

\begin{eqnarray}
\label{HamiltonianCP}
H &=& H_0 + H_I  ,
\nonumber \\
H_0 &=& H_A + H_B + H_F ,
\nonumber \\
H_I  &=& - \bmu_A \cdot \bE (\br_A)  - \bmu_B \cdot \bE (\br_B) ,
\end{eqnarray}
where $H_A$ and $H_B$ are respectively the Hamiltonians of atoms A and B, $\bE (\br )$ indicates the transverse displacement field operator (coinciding, outside the atoms, with the electric field operator, as shown by Equation (\ref{relazD-Ea})), and we have disregarded the term proportional to $\bP_\perp^2 (\br )$, which does not affect interatomic interaction energies. The unperturbed ground state is
$\mid g_A, g_B, \vac \rangle$, where $g_A$ and $g_B$ respectively indicate the ground state of atoms A and B, and $\mid \vac \rangle$ is the photon vacuum state. This state is not an eigenstate of the total Hamiltonian $H$, due to the matter--radiation interaction Hamiltonian. Thus, there is an energy shift of the bare ground state due to the interaction Hamiltonian. Evaluating this energy shift by perturbation theory, it is easily found that the second-order energy shift gives only individual single-atom shifts (the Lamb shift, after mass renormalization), not depending on the distance between the atoms. The first term yielding an interatomic-distance-dependent contribution is at fourth order in the matter--radiation coupling. It is given by
\begin{eqnarray}
\label{fourthorder}
\Delta E_4 &=& -\sum_{I,II,III}\frac {\langle g_A, g_B, \vac \mid H_I \mid III \rangle \langle III \mid H_I \mid II \rangle \langle II \mid H_I \mid I \rangle  \langle I \mid H_I \mid g_A, g_B, \vac \rangle}{(E_I-E_g)(E_{II}-E_g)(E_{III}-E_g)}
\nonumber \\
&\ & +\sum_{I,II} \frac {\mid \langle g_A, g_B, \vac \mid H_I \mid II \rangle \mid^2 \mid  \langle  g_A, g_B, \vac \mid H_I \mid I \rangle \mid^2}{(E_I-E_g)^2(E_{II}-E_g)} ,
\end{eqnarray}
where $I$, $II$ and $III$ indicate atoms-field intermediate states with energies $E_I$, $E_{II}$ and  $E_{III}$ (with the exclusion, in the sums, of the unperturbed state), respectively; $E_g$ is the energy of the unperturbed ground state. The second term in (\ref{fourthorder}) does not contribute to the potential energy, while the first  one gives twelve relevant diagrams, each representing an exchange of two virtual photons between the atoms. Using the Hamiltonian (\ref{HamiltonianCP}) and the expression (\ref{ef}) of the electric field, the following expression of the interaction energy between the two atoms is found from (\ref{fourthorder}) \cite{Craig1998,Salam10}
\begin{eqnarray}
\label{CP1}
\Delta E &=& -\sum_{ps}\sum_{\bk \lambda \bkp \lambda '} \frac {2\pi \hbar ck}V \frac {2\pi \hbar ck'}V (\ekl )_i (\ekl )_j (\eklp )_\ell (\eklp )_m
\mu_{Ai}^{gp} \mu_{A\ell}^{pg} \mu_{Bj}^{gs}\mu_{Bm}^{sg} e^{i(\bk +\bkp )\cdot \br}
\nonumber \\
&\ & \times \frac 1{\hbar^3 c^3}\frac{4(k_{pg}+k_{sg}+k)}{(k_{pg}+k_{sg})(k_{pg}+k)(k_{sg}+k)}\left( \frac 1{k+k'} - \frac 1{k-k'} \right) ,
\end{eqnarray}
where $p$, $s$ indicate generic atomic states, $\br = \br_B -\br_A$ is the distance between the two atoms, \mbox{$\hbar ck_{pg}=E_p-E_g $,} and $i,j,\ell ,m$ are Cartesian components (the Einstein convention of repeated indices has been~used).

After taking the continuum limit $\sum_\bk \rightarrow (V/(2 \pi )^3)\int \! dk k^2\int \! d \Omega$, next steps in the calculation are the polarization sum, done using (\ref{polarizsum}),
and angular integration.

We define the function
\begin{eqnarray}
\label{Ffunction}
G_{ij}(k, \br ) &=& \frac 1{k^3} \left( \delta_{ij} \nabla^2 - \nabla_i \nabla_j \right) \frac {e^{ikr}}r
\nonumber \\
&=& \left[ \left( \delta_{ij} -\hat{r}_i \hat{r}_j \right) \frac 1{kr} +  \left( \delta_{ij} -3\hat{r}_i \hat{r}_j \right) \left( \frac i{k^2r^2} -\frac 1{k^3r^3} \right) \right] e^{ikr} ,
\end{eqnarray}
where the differential operators act on the variable $r$, whose real and imaginary parts are
\begin{eqnarray}
\Re \{G_{ij}(k, \br )\} &=& - V_{ij}(k, \br ) = \frac 1{k^3} \left( \delta_{ij} \nabla^2 - \nabla_i \nabla_j \right) \frac {\cos (kr)}r =
\nonumber \\
&=& \left( \delta_{ij} -\hat{r}_i \hat{r}_j \right) \frac {\cos (kr)}{kr} -  \left( \delta_{ij} -3\hat{r}_i \hat{r}_j \right) \left( \frac {\sin (kr)}{k^2r^2} +\frac {\cos (kr)}{k^3r^3} \right)  ,
\label{ReF} \\
\Im \{G_{ij}(k, \br )\} &=& \frac 1{k^3} \left( \delta_{ij} \nabla^2 - \nabla_i \nabla_j \right) \frac {\sin (kr)}r = \frac 1{4\pi} \int \! d\Omega  \left( \delta_{ij} -\hat{k}_i \hat{k}_j \right) e^{i\bk \cdot \br}
\nonumber \\
&=& \left( \delta_{ij} -\hat{r}_i \hat{r}_j \right) \frac {\sin (kr)}{kr} +  \left( \delta_{ij} -3\hat{r}_i \hat{r}_j \right) \left( \frac {\cos (kr)}{k^2r^2} -\frac {\sin (kr)}{k^3r^3} \right) .
\end{eqnarray}

Using the relations above, after lengthy algebraic calculations, Equation (\ref{CP1}) yields
\begin{eqnarray}
\label{CP2}
\Delta E = &-&\frac 4{\pi \hbar c} \sum_{ps} \mu_{Ai}^{gp} \mu_{A\ell}^{pg} \mu_{Bj}^{gs}\mu_{Bm}^{sg} \frac 1{k_{pg}+k_{sg}}
\nonumber \\
&\ & \times \int_0^\infty \! dk k^6 \frac {k_{pg}+k_{sg}+k}{(k_{pg}+k)(k_{sg}+k)} \Re \{ G_{ \ell m}(k, \br )\} \Im \{G_{i j}(k, \br )\} .
\end{eqnarray}
With further algebraic manipulation, after averaging over the orientations of the atomic dipoles, Equation (\ref{CP2}) can be cast in the following form, in terms of the atoms' dynamical polarizability (\ref{dynpol}) evaluated at the imaginary wavenumber ($k=iu$) \cite{Craig1998,Salam10,Salam08,Shahmoon15}
\begin{equation}
\label{CP3}
\Delta E = -\frac {\hbar c}\pi\int_0^\infty \! \! du \alpha_A(iu) \alpha_B(iu) u^6 e^{-2ur}
\left( \frac 1{u^2r^2}+\frac 2{u^3r^3}+\frac 5{u^4r^4}+\frac 6{u^5r^5}+ \frac 3{u^6r^6} \right) .
\end{equation}

This expression is valid for any interatomic separation $r$ outside the region of wavefunctions overlap, and it depends on the relevant atomic transition wavelengths from the ground state, $\lambda_{rg}=2\pi k_{rg}^{-1}$, included in the atoms' polarizability. Two limiting cases of (\ref{CP3}) are particularly relevant, $r \ll \lambda_{rg}$ (near zone), and $r \gg \lambda_{rg}$ (far zone).

In the near zone, Equation (\ref{CP3}) yields
\begin{equation}
\label{CPnz}
\delta E_{\text{near}} = - \frac 2{3} \sum_{ps} \frac {\mid \bmu_A^{pg}\mid^2 \mid \bmu_B^{sg}\mid^2}{E_{pg}+ E_{sg}} \frac 1{r^6},
\end{equation}
which coincides with the London--van der Waals dispersion interaction scaling as $r^{-6}$ \cite{London1930,Margenau1939}, as obtained in  the minimal coupling scheme including electrostatic interactions only, i.e., neglecting contributions from the (quantum) transverse fields. For this reason, this potential is usually called the {\it nonretarded} London--van der Waals potential energy.

In the far zone, Equation (\ref{CP3}) yields
\begin{equation}
\label{CPfz}
\delta E_{\text{far}} = -\frac {23\hbar c}{4\pi} \alpha_A \alpha_B \frac 1{r^7},
\end{equation}
where $\alpha_{A(B)}=\alpha_{A(B)}(0)$ is the static polarizability of the atoms \cite{CP1948,Babb10}. Equation (\ref{CPfz}) shows that the dispersion energy asymptotically scales with the distance as $r^{-7}$, thus faster than the London potential (\ref{CPnz}). This is usually called the {\it retarded} or Casimir--Polder regime of the dispersion interaction energy, where retardation effects due to the transverse fields, that is, quantum electrodynamical effects, change the distance dependence of the potential energy from $r^{-6}$ to $r^{-7}$, at a distance between the atoms given by the transition wavelength from the ground state to the main excited states. It should be noted that the calculation to obtain the expression (\ref{CP3}) can be simplified by using the effective Hamiltonian introduced in Section \ref{sec:effective Ham} \cite{CP69,PPT98,SPR06}.
The existence of an asymptotic behaviour of the dispersion interaction falling more rapidly than the London $r^{-6}$  dependence, was inferred for a long time from the analysis of some macroscopic properties of colloid solutions \cite{VO99}. In addition, the onset of the Casimir--Polder retarded regime has been recently related to the bond length of a diatomic helium molecule \cite{Przybytek12}. Furthermore, the nonretarded interaction between two single isolated Rydberg atoms, at a distance of a few nanometers each other, has been recently measured through the measurement of the single-atom Rabi frequency \cite{Beguin13}.

We just mention that new features in the distance-dependence of the dispersion energy between ground-state atoms appear if the atoms are in a thermal bath at finite temperature. In such a case, a new length scale, related to the temperature $T$ exists, the thermal length $\rho_\text{therm}= \hbar c/(2\pi k_BT)$, where $k_B$ is the Boltzmann constant. This defines a very long-distance regime, $r \gg \rho_\text{therm}$, where the interaction energy is approximately proportional to $T$ and scales with the distance as $r^{-6}$ \cite{McLachlan63,Boyer75,GW99,Barton01}. This thermal regime, usually at a much larger distance than that of the onset of the retarded regime (far zone), has the same distance-dependence of the nonretarded one.

If one (or both) atoms are in an excited state, the long-distance behaviour of the Casimir--Polder potential is different compared to that of ground-state atoms, asymptotically scaling as $r^{-2}$, due to the possibility of exchange of real photons between the atoms \cite{PT93a,BuhmannII-2012}. From a mathematical point of view, this is a consequence of the presence of a resonant pole in the energy shift. Up to very recently, a~long-standing dispute in the literature has been concerned with whether a spatially oscillating modulation in the interaction energy exists or not: different results (with or without space oscillations), both~mathematically correct but physically different, are obtained according to how the resonant pole is circumvented in the frequency integral. Recent results, also based on a time-dependent approach, have given a strong indication that the distance dependence of the force is monotonic, without spatial oscillations \cite{RPP04,Berman2015,Donaire2015,Milonni2015,Barcellona2016}.
Dispersion interactions, between atoms involving higher-order multipoles \cite{PT96,ST96,Salam2006}, or chiral molecules \cite{Jenkins94,Salam2006a,Barcellona2016a}, have been also investigated in the recent literature.

\section{The Three-Body Casimir--Polder Interaction}
\label{sec:three-body}

\begin{figure}[H]
\centering
\includegraphics[scale=0.9]{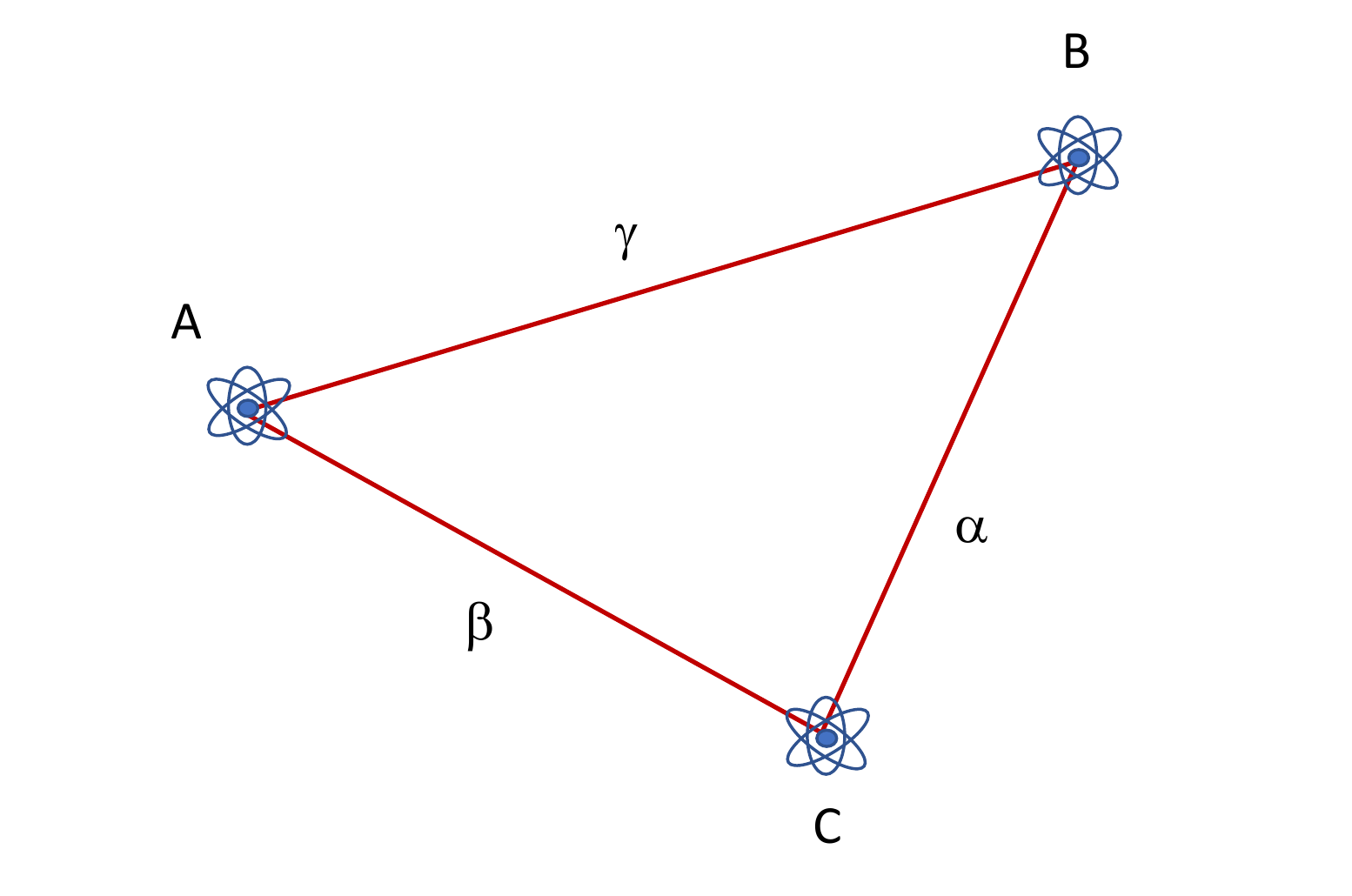}
\caption{Geometrical configuration of the three atoms A, B and C.}
\label{Fig:1}
\end{figure}

Let us consider the case of three atoms, $A$, $B$ and $C$, respectively located at $\br_A$, $\br_B$ and $\br_C$;
their~distances are $\alpha  = \mid \br_C - \br_B \mid$, $\beta  = \mid \br_C - \br_A \mid$, $\gamma  = \mid \br_B - \br_A \mid$, as shown in Figure \ref{Fig:1}.

The Hamiltonian of the system, in the multipolar coupling scheme, is
\begin{equation}
\label{Hamiltonian3}
H = H_A + H_B + H_C+ H_F  - \bmu_A \cdot \bE (\br_A)  - \bmu_B \cdot \bE (\br_B) - \bmu_C \cdot \bE (\br_C) ,
\end{equation}
with a clear meaning of the various terms. The bare ground state is $\mid g_A, g_B, g_C, \vac \rangle$ that, not being an eigenstate of the total Hamiltonian $H$, has an energy shift due to the atoms--field interaction. If this energy shift is evaluated up to sixth order in the interaction, considering only terms depending on the atomic coordinates, and neglecting single-atom energy shifts that do not contribute to interatomic energies, we find
\begin{equation}
\label{EnergyShift6}
\Delta E = \Delta E(A,B) + \Delta E(B,C) + \Delta E(A,C) + \Delta E_3 (A,B,C) ,
\end{equation}
where the first three terms are two-body interaction energies, depending on the coordinates of two atoms only, while $\Delta_3 (A,B,C)$ is a sixth-order non-additive (three-body) contribution containing coordinates and physical parameters of all the three atoms \cite{AZ60,Salam10}. For ground-state atoms, the~three-body term can be expressed in terms of the dynamical polarizabilities of the three atoms or molecules, or polarizable bodies in general.

The effective Hamiltonian (\ref{EffHam}), introduced in Section \ref{sec:effective Ham}, is particularly convenient for the calculation of the three-body interaction between three neutral ground-state atoms, allowing a third-order calculation in place of a sixth-order one, with a considerable reduction of the number of diagrams involved. In our case, the effective Hamiltonian is

\begin{eqnarray}
\label{Hamiltonian3a}
H &=& H_A + H_B + H_C+ H_F  + H_I ,
\nonumber \\
 H_I &=& -\frac 12 \sum_{\bk \lambda} \alpha_A (k) \bE (\bk \lambda , \br_A) \cdot \bE (\br_A)
-\frac 12 \sum_{\bk \lambda} \alpha_B (k) \bE (\bk \lambda , \br_B) \cdot \bE (\br_B)
\nonumber \\
&\ & -\frac 12 \sum_{\bk \lambda} \alpha_C (k) \bE (\bk \lambda , \br_C) \cdot \bE (\br_C) ,
\end{eqnarray}
where $\alpha_i (k)$, with $i=A,B,C$, is the dynamical polarizability of the atoms. Because our effective Hamiltonian is quadratic in the coupling, a third-order perturbation theory allows us to obtain the non-additive term. The general expression of the third-order energy shift is
\begin{eqnarray}
\label{thirdorder}
&\ & \Delta E_3 = -\sum_{I,II} \frac {\langle g_A, g_B, g_C,\vac \mid H_I \mid II \rangle \langle II \mid H_I \mid I \rangle  \langle I \mid H_I \mid g_A, g_B, g_C, \vac \rangle}{(E_I-E_g)(E_{II}-E_g)}
\nonumber \\
&+ & \sum_{I} \frac {\langle g_A, g_B, g_C, \vac \mid H_I \mid g_A, g_B, g_C, \vac \rangle \langle g_A, g_B, g_C, \vac \mid H_I \mid I \rangle \langle I \mid H_I \mid g_A, g_B, g_C, \vac \rangle}{(E_I-E_g)^2} ,
\end{eqnarray}
where $I$ and $II$ are intermediate atoms-field states, different from the unperturbed state.

Application of (\ref{thirdorder}) with the interaction Hamiltonian $H_I$ given by (\ref{Hamiltonian3a}), keeping only terms containing the coordinates of all atoms, after some algebraic calculations, yields \cite{PPT98}
\begin{equation}
\label{3bodyint}
\Delta E_3 (A,B,C) = -\frac {\hbar c}\pi F_{ij}^\alpha F_{j\ell}^\beta F_{\ell i}^\gamma \frac 1{\alpha \beta \gamma}
\int_0^\infty \! du \alpha_A(iu) \alpha_B(iu) \alpha_C(iu) e^{-u(\alpha +\beta +\gamma )} ,
\end{equation}
where, in order to shorten notations, we have defined the differential operator
\begin{equation}
\label{Fdiffop}
F_{mn}^r = \left( -\delta_{mn} \nabla^2 + \nabla_m \nabla_n \right)^r
\end{equation}
(the superscript indicates the space variable upon which the differential operators act). The expression (\ref{3bodyint}), here obtained by a simpler third-order perturbative approach using the effective Hamiltonian (\ref{Hamiltonian3a}), coincides with the standard sixth-order result obtained from the multipolar Hamiltonian \cite{AZ60}.

In the far zone (retarded regime), only low-frequency field modes give a relevant contributions, and we can replace the dynamical polarizabilities with the static ones. The integral over $u$ is then straightforward, and we get
\begin{eqnarray}
\label{3bodyintfz}
\Delta E_3 (A,B,C) \simeq -\frac {\hbar c}\pi \alpha_A \alpha_B \alpha_C
F_{ij}^\alpha F_{j\ell}^\beta F_{\ell i}^\gamma
\frac 1{\alpha \beta \gamma (\alpha + \beta + \gamma )} ,
\end{eqnarray}
where $\alpha_i$ $(i=A,B,C)$ are the atomic static polarizabilities.

In the specific case of three atoms in the far zone of each other, with an equilateral triangle configuration of side $r$ ($\alpha = \beta = \gamma = r$), Equation (\ref{3bodyintfz}) gives \cite{PT85}
\begin{equation}
\label{3bodyintfzeq}
\Delta_3(A,B,C) \simeq \frac {2^4 \cdot 79}{3^5} \frac {\hbar c}\pi \frac {\alpha_A \alpha_B \alpha_C}{r^{10}} .
\end{equation}

In the same geometrical configuration, Equation (\ref{3bodyint}) gives the nonretarded result (near zone) scaling as $r^{-9}$ \cite{AT43}. The $r^{-10}$ and $r^{-9}$ distance scaling of the three-body-component should be compared with the $r^{-7}$ and $r^{-6}$ scaling of the two-body component, respectively in the far and near zone. While the two-body interaction is always attractive, the three-body component can be attractive or repulsive according to the geometrical configuration of the three atoms \cite{PT85}. Three-body interactions between molecules involving higher-order multipoles have been also considered in the literature~\cite{Salam13,Salam14,BS18}.

\section{Two- and Three-Body Dispersion Interactions as a Consequence of Vacuum Field Fluctuations}
\label{sec:vf}

An important point when dealing with van der Waals and Casimir--Polder dispersion interactions is the formulation of physical models aiming to explaining their origin, stressing quantum aspects, and giving physical insights of the processes involved. This is important from two points of view: they~allow for understanding what the basic origin of these interactions is, highlighting quantum and classical aspects; they can provide useful computational methods to simplify their evaluation, in particular in more complicated situations such as many-body interactions, or when boundary conditions are present, as well as when the atoms are in a noninertial motion.

In this section, we describe physical models for dispersion interactions based on the existence and properties of vacuum field fluctuations. These models clearly show how van der Waals and Casimir--Polder interactions between atoms, and polarisable bodies in general, in different conditions, can be attributed to the existence of the zero-point energy and vacuum fluctuations, specifically their change due to the presence of matter, and/or their spatial correlations. At the end of this review, we will briefly address the fundamental and conceptually subtle question if all this {\it proves} the real existence of vacuum fluctuations. In fact, although the results discussed in the following give strong support to the real existence of the zero-point energy, they are not a definitive confirmation of its existence because all of these effects can in principle be obtained also from considerations based on source fields, without reference to vacuum fluctuations \cite{Milonni82,Milonni88,Milonni1994}.

We now introduce and apply two different models, based on the properties of vacuum field fluctuations and {\it dressed} vacuum field fluctuations, to obtain on a physically transparent basis the two- and three-body dispersion interactions between polarizable bodies, even in the presence of a boundary such as a reflecting mirror. The first method is based on vacuum field energy densities, and~the second on vacuum field spatial correlations. These models give physical insights on the origin of the ubiquitous dispersion interactions, and also provide useful tools to calculate them in more complicated situations, which we will apply in Section \ref{sec:boundaries} when a reflecting mirror close to the atoms is~present.

\subsection{Dressed Field Energy Densities}
\label{sec:endens}

Let us consider two or more atoms, or in general polarizable bodies, in the unbounded space, at~zero temperature, interacting with the quantum electromagnetic field in the vacuum state. The~bare ground state $\mid g \rangle$, consisting of the ground state of the atoms and the vacuum state of the field, is~not an eigenstate of the total Hamiltonian. The presence of matter perturbs the field \cite{Power1966}, due to continuous emission and absorption of virtual photons \cite{CPP83,PCP85}, yielding the dressed ground state $\mid \gd \rangle$ of the system. In such a state, the atoms are {\it dressed} by the virtual photons emitted and reabsorbed by them.  The presence of this {\it cloud} of virtual photons determines a change of the fluctuations of the electric and magnetic field around the atoms or, equivalently, of the electric and magnetic energy densities; the {\it renormalized} energy densities (that is, after subtraction of the spatially uniform bare field energy densities, present even in absence of the atom) are given by
\begin{eqnarray}
\langle \gd \mid {\cal H}_{\text{el}}(\br ) \mid \gd \rangle_\text{ren} &=& \frac 1{8\pi} \langle \gd \mid \bE^2(\br ) \mid \gd \rangle - \frac 1{8\pi} \langle g \mid \bE^2(\br ) \mid g \rangle ,
\label{elendens1} \\
\langle \gd \mid {\cal H}_{\text{mag}}(\br ) \mid \gd \rangle_\text{ren} &=& \frac 1{8\pi} \langle \gd \mid \bB^2(\br ) \mid \gd \rangle - \frac 1{8\pi} \langle g \mid \bB^2(\br ) \mid g \rangle ,
\label{magendens1}
\end{eqnarray}
where $\mid g \rangle$ and $\mid \gd \rangle$ are respectively the bare and the dressed ground state of the system.

Let us first consider just one ground-state atom, $A$, located at $\br_A=0$. The dressed ground state can be obtained by perturbation theory; it is convenient to use the multipolar coupling Hamiltonian obtained in Section \ref{sec:Hamiltonian}, or the effective Hamiltonian (\ref{EffHam}), because in such a case we will get the transverse displacement field that, according to (\ref{relazD-Ea}), coincides with the total (longitudinal plus transverse) electric field $\bE (\br )$ for $\br \neq 0$ (this is an essential point because it is the total electric field, and not just its transverse part, which is involved in the interaction with the other atoms). Thus, the energy density we will obtain is that associated with the total electric field and to the magnetic field. Applying first-order perturbation theory to the effective Hamiltonian (\ref{EffHam}), the dressed ground state at first order in the polarizability, that is, at second order in the electric charge, is
\begin{equation}
\label{drgrstate}
\mid \gd \rangle = \mid g_A \vac \rangle - \frac \pi V \sum_{\bk \lambda \bkp \lambda '} \alpha_A (k) \ekl \cdot \eklp \frac {kk'}{k+k'}\mid g_A, 1_{\bk \lambda} 1_{\bkp \lambda '} \rangle ,
\end{equation}
showing admixture with two virtual photons. We now evaluate the average value of the electric and magnetic energy density over the state (\ref{drgrstate}). At first order in $\alpha_A(k),$ we get \cite{PRS13}
\begin{eqnarray}
 \langle \gd \mid {\cal H}_{\text{el}}(\br ) \mid \gd \rangle_\text{ren} &=& \frac{\hbar c}{4\pi^3}\int_0^\infty \! dk  \int_0^\infty \! dk'  \alpha_A (k) \frac {k^3k'^3}{k+k'} \Big[ j_0(kr)j_0(k'r)
 \nonumber \\
 &\ & -\frac 1{k'r} j_0(kr) j_1(k'r) -\frac 1{kr}j_0(k'r) j_1(kr)+ \frac 3{kk'r^2}  j_1(kr) j_1(k'r) \Big] ,
 \label{elendens2}\\
 \langle \gd \mid {\cal H}_{\text{mag}}(\br ) \mid \gd \rangle_\text{ren}  &=& - \frac{\hbar c}{2\pi^3}\int_0^\infty \! dk  \int_0^\infty \! dk'  \alpha_A (k) \frac {k^3k'^3}{k+k'} j_1(kr)j_1(k'r) ,
 \label{magendens2}
\end{eqnarray}
where $j_n(x)$ are spherical Bessel functions \cite{NIST10}. The integrals over $k$ and $k'$ in (\ref{elendens2}) and (\ref{magendens2}) can be decoupled by using the relation $(k+k')^{-1}=\int_0^\infty \! d\eta \exp (-(k+k')\eta )$. Explicit evaluation \mbox{of (\ref{elendens2}) and
(\ref{magendens2})} yields a positive value for the renormalized electric electric energy density, and a negative value for the magnetic one. This means that the presence of an electric dipolar source increases the electric vacuum energy density, while it decreases the magnetic vacuum energy density, with respect to their value in the absence of the source (an analogous consideration holds for the field fluctuations, of course). This is a microscopic analogue of what happens in the case of the plane conducting boundary discussed in Section \ref{sec:vacuum}, as Equations (\ref{elffplate}) and (\ref{magffplate}) show.

Two limiting cases of (\ref{elendens2}) and (\ref{magendens2}) are particularly relevant, according to the distance $r$ from the atom in comparison with a relevant atomic transition wavelength $\lambda_0$ from the ground state, similarly to the dispersion energy between two ground-state atoms. In the {\it far zone}, $r \gg \lambda_0$, it is easy to show that only small-frequency field modes significantly contribute to the $k$, $k'$ integrals in Equations (\ref{elendens2},\ref{magendens2}), and we can approximate the (electric) dynamic polarizability with the static electric one, $\alpha_A(0) = \alpha_A^E$. We thus obtain
\begin{equation}
\label{endenfar}
 \langle \gd \mid {\cal H}_{\text{el}}(\br ) \mid \gd \rangle_\text{ren} \simeq \frac {23\hbar c \alpha_A^E}{(4\pi )^2} \frac 1{r^7}; \, \, \, \, \,  \langle \gd \mid {\cal H}_{\text{mag}}(\br ) \mid \gd \rangle_\text{ren} \simeq -\frac {7\hbar c \alpha_A^E}{(4\pi )^2} \frac 1{r^7} .
\end{equation}

In the {\it near zone}, $r \ll \lambda_0$, we find that the renormalized electric energy density scales as $1/r^6$, and the magnetic one as $1/r^5$ \cite{PP87a,CPPP95a}. In this limit, the electric energy density is essentially due to the electrostatic (longitudinal) field, as an explicit calculation in the minimal coupling scheme, where longitudinal and transverse contributions are separated (see Section \ref{sec:Hamiltonian}), shows \cite{PP87a}.

The results above clearly point out how the presence of a field source, a dipolar atom in our case, changes zero-point energy densities and field fluctuations in the space around it, and, more in general, that matter can influence the properties of the quantum vacuum. This is the microscopic analogue of the change of the vacuum energy densities due to a boundary discussed in Section \ref{sec:vacuum}, or in the two-plate case of the Casimir effect \cite{Casimir48}; the essential difference is that in the present case, matter is described through its Hamiltonian, with its internal degrees of freedom, while in the case of macroscopic boundaries its presence is introduced by a (classical) boundary condition on the field~operators.

In the far zone, the change of the zero-point energy due to the presence of atom A can be probed through a second atom, B, considered a polarizable body with static electric polarizability $\alpha_B^E$ and located at $\br_B$. Its interaction energy with field fluctuations can be written as $\Delta E = -\alpha_B^E\langle E^2(\br_B)\rangle/2$, as indicated by the existence of an effective Hamiltonian of the form (\ref{EffHamSt}); this is also supported by the classical expression of the interaction energy of an electric field with an induced electric dipole moment. Similarly, in the far zone, the magnetic energy density can be probed through a magnetically polarizable body with static magnetic polarizability $\alpha_B^M$. We stress that this is rigorously valid only in the far zone because, in the near zone, where the contribution of high-frequency photons is relevant \cite{PP87a}, the~{\it test} atom B responds differently to each Fourier component of the field, according to its dynamical polarizability $\alpha_B(k)$ (see Equation (\ref{EffHam})). The distance-dependent part of the interaction energy is then
\begin{eqnarray}
\Delta E^{EE}(\br ) &=& -\frac 12 \alpha_B^E   \langle \gd \mid E^2(\br_B ) \mid \gd \rangle_\text{ren} = - \frac {23 \hbar c}{4\pi} \alpha_A^E \alpha_B^E \frac 1{r^7} ,
\label{CP-EE} \\
\Delta E^{EM}(\br ) &=& -\frac 12 \alpha_B^M   \langle \gd \mid B^2(\br_B ) \mid \gd \rangle_\text{ren} =  \frac {7 \hbar c}{4\pi} \alpha_A^E \alpha_B^M \frac 1{r^7}  ,
\label{CP-EM}
\end{eqnarray}
where $r$ is the distance between the two atoms, and the subscript {\em ren} indicates that spatially uniform bare terms have been subtracted. These expressions coincide with the dispersion interaction energy in the far-zone as obtained by a fourth-order perturbative calculation, as discussed in Section \ref{sec:vdW-CP} \mbox{(see, for example,} \cite{FS70,PP87a}); from the signs of (\ref{CP-EE}) and (\ref{CP-EM}), it should be noted that the electric-electric dispersion force is attractive (because the presence of the electric dipolar atom decreases the electric energy density), and the electric-magnetic dispersion force is repulsive (because the presence of the electric dipolar atom increases the magnetic energy density).

This approach shows the strict relation between (renormalized) electric and magnetic vacuum energy densities and the far-zone dispersion interaction energy; moreover, it allows considerable simplification of the calculations, that in this way is split in two independent steps: evaluation of the renormalized energy densities of the field first, and then the evaluation of their interaction with the other electrically or magnetically polarizable body. This finding, besides giving a sharp insight on the origin of the retarded dispersion interaction, can be particularly useful in more complicated cases such as many-body dispersion interactions, or in the presence of boundary conditions (see Section \ref{sec:boundaries}).

Let us now consider the dispersion interaction between three ground-state atoms, A, B and C, respectively located at positions $\br_A$, $\br_B$ and $\br_C$. We concentrate on the three-body component of their interaction, discussed in Section \ref{sec:three-body}. We will show that, analogously to the case of the two-body potential, the physical origin of the retarded (far zone) three-body component can be clearly attributed to vacuum fluctuations and field energy densities, as modified by the presence of the atoms \cite{PP99}. The~Hamiltonian of our system is given by (\ref{Hamiltonian3a}); because we are limiting our considerations to the far zone only, the~interaction term can be more simply expressed in terms of the Craig--Power effective Hamiltonian (\ref{EffHamSt}); thus, we have
\begin{equation}
\label{CPH3}
H_I = -\frac 12 \sum_{i=A,B,C} \alpha_i \bE^2 (\br_i)
\end{equation}
(this expression can be straightforwardly obtained from (\ref{Hamiltonian3a}) by replacing the dynamical polarizabilities with the corresponding static ones).
The first step is to obtain the dressed ground state of the pair of atoms A and B, up to second order in the polarizabilities, i.e., up to the fourth order in the electronic charge. Writing down only terms relevant for the subsequent calculation of the three-body interaction, that is, only terms that at the end of the calculation yield contributions containing the polarizabilities of all atoms, we have
\begin{eqnarray}
\label{gdAB}
\mid \tilde{g}_{AB} \rangle &=& \mid g_A g_B, \vac \rangle -
\left( \frac \pi V \alpha_A \sum_{\bk \lambda \bkp \lambda '} \left( \ekl  \cdot \eklp \right) \frac {\sqrt{\wk \wkp}}{\wk + \wkp} e^{-i(\bk +\bkp )\cdot \br_A }  \mid g_A g_B, \bk \lambda \bkp \lambda ' \rangle \right.
\nonumber \\
&\ &  + (A \rightarrow B) \Bigg) - \Bigg( \frac {8\pi^2}{V^2} \alpha_A \alpha_B \sum_{\bk \lambda \bkp \lambda '} \left( \ekl  \cdot \eklp \right)
\left( \eklp \cdot \eklpp \right)
\frac {\wkpp \sqrt{\wk \wkp}}{(\wkpp +\wkp)(\wk +\wkp )}
\nonumber \\
&\ & \ \ \ \ \ \ \  \times  e^{-i(\bkp +\bkpp)\cdot \br_A} e^{-i(\bk -\bkpp)\cdot \br_B} \mid g_A g_B, \bk \lambda \bkp \lambda ' \rangle + (\br_A \leftrightarrow \br_B)
\Bigg) .
\end{eqnarray}

We now evaluate the average value of the electric field fluctuations at the position $\br_C$ of atom C on the dressed state (\ref{gdAB}), which is the quantity
$\langle \tilde{g}_{AB} \mid E^2(\br_C) \mid \tilde{g}_{AB} \rangle$. After multiplying times $-\alpha_C/2$, it~gives the interaction energy of atom C with the (renormalized) field fluctuations generated by atoms A and B. This quantity contains a term proportional to the atomic polarizability of the three atoms, yielding the three-body component of the dispersion interaction. After lengthy algebraic calculations, we~finally~obtain
\begin{equation}
\label{intener3}
-\frac 12 \alpha_C \langle \tilde{g}_{AB} \mid E^2(\br_C) \mid \tilde{g}_{AB} \rangle = -\frac {\hbar c}\pi \alpha_A \alpha_B \alpha_C
F_{ij}^\alpha F_{j\ell}^\beta F_{\ell i}^\gamma
\frac 1{\alpha \beta \gamma (\alpha + \beta + \gamma )} \, ,
\end{equation}
where the distances $\alpha$, $\beta$ and $\gamma$ have been defined in Section \ref{sec:three-body}, and the superscripts indicate the variables the differential operators act on. Two analogous expressions are obtained by exchanging the role of the atoms. The total interaction energy is then obtained by averaging on these three (equal) contributions: the same expression obtained by a direct application of perturbation theory, given by (\ref{3bodyint}), is thus obtained \cite{PP99}.
This shows that the three-body component of the retarded dispersion interaction between three atoms (far zone, Casimir--Polder regime) at sixth order in the electron charge, can be directly and more easily obtained as the interaction energy of one atom with the renormalized electric field fluctuations generated by the other two atoms, that, at the fourth order in the charge, show a sort of interference effect. In other words, at the fourth order, the fluctuations in the presence of two atoms are not simply the sum of those due to the individual atoms \cite{CPP87}. This approach, similarly to the two-body case previously analyzed, has two remarkable features: firstly, it gives a physically transparent insight on the origin of three-body dispersion interactions, particularly significant in the retarded regime; secondly, it allows a considerable simplification of the calculation, which is separated in two successive steps: evaluation of the renormalized field energy density due to the presence of the ``source'' atoms A and B, and successive interaction of them with the third atom (C).

The approach outlined in this section can be directly applied also to the retarded Casimir--Polder interaction between an atom and a perfectly reflecting wall. Renormalized electric vacuum fluctuations are given by (\ref{elffplate}) and (\ref{magffplate}). If an electrically polarizible body A, such as an atom or a molecule, is placed at a distance $z$ from the conducting wall at $z=0$, the second-order interaction energy in the far zone is given by
\begin{equation}
\label{atom-wall}
\Delta E_\text{atom-wall}(z) = -\frac 12 \alpha_A \langle \vac \mid E^2(z) \mid \vac \rangle = -\frac {3\hbar c}{8\pi} \alpha_A \frac 1{z^4} ,
\end{equation}
where $\alpha_A$ is the electric static polarizability of the atom. In a quasi-static approach (as that we are assuming for the dispersion force between atoms, too), the Casimir--Polder force on the atom is given~by
\begin{equation}
\label{atom-wall-force}
F_\text{atom-wall}(z)=- \frac d{dz}\left( \Delta E_\text{atom-wall}(z) \right) = -\frac {3\hbar c}{2\pi} \alpha_A \frac 1{z^5},
\end{equation}
where the minus sign implies that the force is attractive \cite{CP1948,HS91,Messina08,Barton74}. Similar considerations allow us to evaluate the far-zone magnetic interaction, starting from (\ref{magffplate}). Nonperturbative methods to evaluate the atom-wall Casimir--Polder interaction, based on Bogoliubov-type transformations and modelling the atoms as harmonic oscillators, have been also proposed in the recent literature \cite{Passante12,CKP05}.

\subsection{Vacuum Field Correlations}
\label{vacfieldcorr}

A different approach showing the deep relation between dispersion interactions and zero-point field fluctuations involves the spatial correlation function of vacuum fluctuations, which, in the unbounded vacuum space is given by (\ref{correl-mode}), for a single field mode, and by (\ref{ef-corr}) for the field operator. Let~us assume considering two isotropic ground-state atoms or molecules, A and B, respectively located at $\br_A$ and $\br_B$. Zero-point fluctuations induce instantaneous dipole moments in the two atoms according~to
\begin{equation}
\label{inddip}
\mu_{\ell i}^{\text{ind}}(\bk \lambda ) = \alpha_\ell (k) E_i (\bk \lambda , \br_\ell ) ,
\end{equation}
where $\mu_{\ell i}^{\text{ind}}(\bk \lambda )$ is the $(\bk \lambda )$ Fourier component of the $i$ cartesian component of the induced dipole moment $\bmu_\ell$ of atom $\ell = A,B$, and $\alpha_\ell (k)$ is the isotropic dynamical polarizability of atom $\ell$. Because, \mbox{according to  (\ref{correl-mode}),} vacuum fluctuations are space-correlated, the induced dipole moments in atoms A and B will be correlated too. Their dipole--dipole interaction then yields an energy shift given by \cite{PT93}
\begin{equation}
\label{enshiftcorr}
\Delta E = \sum_{\bk \lambda} \langle \mu_{A i}^{\text{ind}}(\bk \lambda ) \mu_{B j}^{\text{ind}}(\bk \lambda ) \rangle V_{ij}(k,\br =\br_B-\br_A) ,
\end{equation}
where $V_{ij}(k,\br )$ is the potential tensor for oscillating dipoles at frequency $ck$, whose expression is given by (\ref{ReF}) \cite{BornWolf,MLP64}. Substitution of (\ref{inddip}) into (\ref{enshiftcorr}) yields
\begin{equation}
\label{enshiftcorr2}
\Delta E = \sum_{\bk \lambda} \alpha_A(k) \alpha_B(k) \langle \vac \mid E_i (\bk \lambda , \br_A ) E_i (\bk \lambda , \br_B )\mid \vac \rangle V_{ij}(k,\br ) ,
\end{equation}
where $\br = \br_B -\br_A$ is the distance between the two atoms. In (\ref{enshiftcorr2}), the vacuum spatial correlation for the electric field explicitly appears. After some algebraic calculation involving polarization sum and angular integration, evaluation of  (\ref{enshiftcorr2}) gives, both in the nonretarded and retarded regime, \mbox{the result (\ref{CP3})} of Section \ref{sec:vdW-CP}, as obtained by a fourth-order perturbation theory, or by a second-order perturbation theory if the effective Hamiltonian (\ref{EffHam}) is used \cite{PT93,PPR03}.

In this approach, the origin of dispersion interaction is evident from a physical point of view: vacuum fluctuations, being spatially correlated, induce correlated dipole moments in the two atoms, which then interact with each other through a (classical) oscillating dipolar term. In this approach, the only quantum aspect of the electromagnetic field is the existence of the spatially correlated vacuum fluctuations. All other aspects involved are just classical ones. The interaction between the two induced dipoles, expressed by $ V_{ij}(k,\br )$, has, in fact, an essentially classical origin. The $r^{-7}$ behaviour in the retarded Casimir--Polder regime (far zone) can be now easily understood: the spatial field correlation function behaves as $r^{-4}$, as Equation (\ref{ef-corr}) shows, and thus the correlation function of the induced dipole moments has the same space dependence. By taking into account that the (classical) interaction energy between the induced dipoles depends as $r^{-3}$ from $r$, the $r^{-7}$ far-zone interaction energy is immediately obtained. In the near zone, the situation is a bit different and somehow more complicated because the atoms respond to the field fluctuations differently for any Fourier component, according to their frequency-dependent dynamical polarizability.

This approach, based on spatial field correlations, can be generalized to three-body van der Waals and Casimir--Polder interactions; in such case, the key element are the {\it dressed} vacuum fluctuations \cite{CP96,CP97,CMPR01}. We consider three isotropic ground-state atoms or molecules (or polarizable bodies), A, B, and C, respectively located at $\br_A$, $\br_B$ and $\br_C$. In order to generalize the method outlined before to the three-body interaction, we first consider how the presence of one of the three atoms, A, changes the spatial correlation function of the electric field, yielding dressed field correlations. The multipolar-coupling interaction Hamiltonian of atom A is
\begin{equation}
\label{intHam3}
H_I^A = - \bmu_A \cdot \bE (\br_A) .
\end{equation}

The second-order dressed ground state of atom A (as atoms B and C were absent) is
\begin{eqnarray}
\label{dgsA}
\mid \tilde{g}_A\rangle &=& N \mid g_A \vac \rangle -\frac 1{\hbar c}\sum_m \sum_{\bk \lambda}
\frac {\langle m, \bk \lambda \mid H_I^A \mid g_A, \vac \rangle}{k+k_{mg}} \mid m, \bk \lambda \rangle
\nonumber \\
&\ & + \frac 1{(\hbar c)^2}\sum_{mn}\sum_{\bk \lambda \bkp \lambda '}
\frac {\langle n, \bk \lambda \bkp \lambda '\mid H_I^A\mid m, \bk \lambda \rangle \langle m, \bk \lambda \mid H_I^A \mid g_A, \vac \rangle}{(k+k_{mg})(k+k'+k_{ng})} \mid n,  \bk \lambda \bkp \lambda ' \rangle ,
\end{eqnarray}
where $N$ is a normalization factor, $\hbar ck_{mg}=E_m-E_g$, and $m,n$ are basis states of atom A. This state has virtual admixtures with one- and two-photon states, which, in the space around atom A, modify vacuum fluctuations, specifically their spatial correlations. We can now evaluate the average value of the electric-field correlation function on the dressed ground state (\ref{dgsA}), at the positions $\br_B$ and $\br_C$ of atoms B and C, respectively. It can be expressed in the form
\begin{equation}
\label{dscf}
\langle \tilde{g}_A \mid E_i (\bk \lambda ,\br_B) E_j (\bkp \lambda ',\br_C) \mid  \tilde{g}_A \rangle = \langle E_i (\bk \lambda ,\br_B) E_j (\bkp \lambda ',\br_C) \rangle_0 + \langle E_i (\bk \lambda ,\br_B) E_j (\bkp \lambda ',\br_C) \rangle_A^{\text{gs}}  .
\end{equation}

The first term on the right-hand-side of (\ref{dscf}) is the same of the bare vacuum, as given by Equation~(\ref{correl-mode}), and it is independent from the position of atom A; thus, it does not contribute to the three-body interaction, and we disregard it. The second term is the second-order correction to the field correlation due to the presence of the ground-state atom A, and contains correlation between different modes of the field; its explicit expression is  \cite{CP97}
\begin{eqnarray}
\label{cfgs}
&\ & \langle E_i (\bk \lambda ,\br_B) E_j (\bkp \lambda ',\br_C) \rangle_A^{\text{gs}}= \frac {4\pi^2kk'}{V^2} \sum_m (\bmu_A^{gm} \cdot \eklp ) (\bmu_A^{mg} \cdot \ekl )
(\ekl)_i  (\eklp)_j
\nonumber \\
&\ & \times \Big\{ \frac {e^{-i\bk \cdot (\br_B -\br_A)} e^{i\bkp \cdot (\br_C -\br_A)}}{(k+k_{mg})(k'+k_{mg})}
+ \frac {e^{i\bk \cdot (\br_B -\br_A)} e^{i\bkp \cdot (\br_C -\br_A)}}{k+k'} \left( \frac 1{k+k_{mg}} +\frac 1{k'+k_{mg}} \right) +\text{c.c.} \Big\} .
\end{eqnarray}
After some algebraic manipulation, this expression can be also cast in terms of the dynamic polarizability of atom A.

The dressed correlated vacuum field induces and correlates dipole moments in atoms B and C according to (\ref{inddip}), and this gives an interaction energy between B and C that, for the part depending from the presence of A (see Equation (\ref{cfgs})), is
\begin{equation}
\label{DeltaEBC}
\Delta E_{BC} = \sum_{\bk \lambda \bkp \lambda '}\alpha_B(k) \alpha_C(k') \langle E_i (\bk \lambda ,\br_B) E_j (\bkp \lambda ',\br_C) \rangle_A^{\text{gs}} V_{ij}(k \br_B, k' \br_C) ,
\end{equation}
where $V_{ij}(k \br_B, k' \br_C)$ is an appropriate generalization, to the present case of dipoles oscillating at frequencies $ck$ and $ck'$, of the potential tensor previously introduced for the two-body potential. We~assume the symmetric expression

\begin{equation}
\label{potsimm}
V_{ij}(k \br_B, k' \br_C) = \frac 12 \left( V_{ij}(k,\br_B) + V_{ij}(k',\br_C) \right) .
\end{equation}

The quantity $\Delta E_{BC}$ in (\ref{DeltaEBC}) is the interaction between B and C in the presence of A. If we symmetrize on the role of the three atoms, after some algebraic calculations, we obtain the correct expression of the three-body potential energy in the far zone,
\begin{eqnarray}
\label{DeltaEABC}
\Delta E_{ABC}^\text{gs} &=& \frac 23  \left( \Delta E_{AB} + \Delta E_{BC} + \Delta E_{AC} \right)
\nonumber \\
&=&  -\frac {\hbar c}\pi \alpha_A \alpha_B \alpha_C
F_{ij}^\alpha F_{j\ell}^\beta F_{\ell i}^\gamma \frac 1{\alpha \beta \gamma (\alpha + \beta + \gamma )} ,
\end{eqnarray}
where the distances $\alpha$, $\beta$ and $\gamma$ between the atoms have been defined in Section \ref{sec:three-body} (see also Figure \ref{Fig:1}). This~expression coincides with the interaction energy $\Delta E_3 (A,B,C)$ of Equation (\ref{3bodyintfz}), originally obtained through a sixth-order perturbative calculation based on the multipolar Hamiltonian \cite{AZ60}, or a third-order calculation using the effective Hamiltonian (\ref{EffHam}) \cite{PPT98}. This gives a new physical picture of the three-body interaction: it originates from the interaction of two atoms (B and C), whose dipole moments have been induced and correlated by the spatially correlated electric field fluctuations, as modified ({\it dressed}) by atom A.

The physical picture of the three-body dispersion interaction for ground-state atoms outlined above, based on dressed vacuum field correlations, can be extended to atoms in excited states too, both in static and dynamical (time-dependent) situations. Let us consider one excited atom (A), approximated as a two-level system with frequency $\w0 =ck_0$, and two ground-state atoms (B and C), interacting with the quantum electromagnetic field in the vacuum state. We proceed similarly to the case of three ground-state atoms, taking the atom A as a sort of source atom; firstly, we evaluate its dressed ground state, and then the dressed spatial correlation on this state of the electric field at the position of the other two atoms. We use the multipolar coupling Hamiltonian (we cannot use the effective Hamiltonian (\ref{EffHam}) for the excited atom A because it is not valid on the energy shell, and we are now dealing with an excited state). The main difference, with respect to the previous case of a ground-state atom, is that a pole at $ck_0$ is now present in the frequency integration, and this yields an extra (resonant) term in the correlation function. After some algebra, we obtain \cite{PPR05}
\begin{eqnarray}
\label{corrfuntexc}
\langle E_i (\bk \lambda ,\br_B) E_j (\bkp \lambda ',\br_C) \rangle_A^{\text{es}} &=&
-\frac {2k_0 \mu_\ell^A \mu_m^A}\pi
F_{i\ell}^\gamma F_{jm}^\beta
 \frac 1{\beta \gamma}
\int_0^\infty \! \! du \frac {e^{-u(\beta + \gamma )}}{k_0^2+u^2}
\nonumber \\
&\ & + 2 \mu_\ell^A \mu_m^A F_{i\ell}^\gamma F_{\ell j}^\beta \frac {\cos (k_0(\gamma -\beta ))}{\beta \gamma} ,
\end{eqnarray}
where the bare term, independent of the position of atom A, has been disregarded because it does not contribute to the three-body interaction (it only contributes to the two-body components, as previously shown in this section). In the first term of (\ref{corrfuntexc}), we recognize the quantity
$ -2k_0 \mu_\ell^A \mu_m^A/(\hbar c (k_0^2+u^2))=\alpha_A^\text{es}(iu)$
as the dynamical polarizability for the excited state of the two-level atom A, evaluated at imaginary frequencies. It is easy to see that this term is the same of the ground-state case (\mbox{see Equation (\ref{cfgs})}), except for the presence of the excited-state polarizability of atom A. The second term is new, and originates from the presence of a resonant pole at $k=k_0$ in the integration over $k$; it~contains contributions from only the frequency $ck_0$. With a procedure analogue to that leading to~(\ref{DeltaEABC}) for ground-state atoms, including appropriate symmetrization over the atoms, for the three-body interaction energy of one excited- and two ground-state atoms, we finally obtain \cite{PPR05},
\begin{eqnarray}
\label{DeltaEABC2}
\delta E_{ABC}^\text{es} &=& -\mu_n^A \mu_p^A \alpha_B(k_0) \alpha_C(k_0) F_{ij}^\alpha F_{j\ell}^\beta F_{\ell i}^\gamma \frac 1{\alpha \beta \gamma}
\left[ \cos (k_0(\beta -\gamma +\alpha )) + \cos (k_0(\beta -\gamma -\alpha )) \right]
\nonumber \\
&\ &  -\frac {\hbar c}\pi \alpha_A^\text{es} \alpha_B \alpha_C
F_{ij}^\alpha F_{j\ell}^\beta F_{\ell i}^\gamma \frac 1{\alpha \beta \gamma}  \int_0^\infty \! du \alpha_A^\text{es}(iu) \alpha_B(iu) \alpha_C(iu) e^{-u(\alpha +\beta +\gamma )} .
\end{eqnarray}

The second term in (\ref{DeltaEABC2}) has the same structure of the ground-state three-body dispersion energy discussed previously, except for the presence of the excited-state polarizability of atom A in place of the ground-state polarizability. The first term is a new one, arising from the resonant photon exchange, characterized by a different scaling with the distance and by the presence of space oscillations with frequency $ck_0$, the transition frequency of the excited two-level atom \cite{PPR05,PT95,PT95a}. Although this result, analogously to (\ref{DeltaEABC}), can be also obtained by a perturbative approach, the derivation outlined in this section gives a clear and transparent physical insight of the three-body interaction as due to the modification of zero-point field correlations due to the presence of one atom, or in general a polarizable body.
It should be noted that, in the present case of one excited atom, we have neglected its spontaneous decay, and thus are assuming that the times considered are shorter than its decay time.
The method used can be usefully extended also to dynamical (i.e., time-dependent) situations, in~order to investigate dynamical many-body dispersion (van der Waals and Casimir--Polder) interactions between atoms \cite{PPR06,PPR07,RPP07}, as well as dynamical atom--surface interactions \cite{VP08,SHP03,HRS04,MVP10,Messina14,Haakh14,Armata16}, for example during the dynamical dressing of one atom starting from a nonequilibrium situation.

\section{Casimir--Polder Forces between Atoms Nearby Macroscopic Boundaries}
\label{sec:boundaries}

In this section, we show that the methods discussed in the previous section, based on the properties of the quantum electrodynamical vacuum, can be extended to dispersion interactions between atoms or molecules in the presence of a macroscopic body, specifically an infinite plate of a perfectly reflecting material. This is relevant because it shows that dispersion interactions can be manipulated (enhanced or inhibited) through the environment---for example, a cavity \cite{PT82} or a metallic waveguide \cite{HS15,Dung16}, similarly also to other radiative procsses such as, for example, the radiative energy transfer \cite{Weeraddana17,Fiscelli18}.

\begin{figure}[H]
\centering
\includegraphics[scale=0.5]{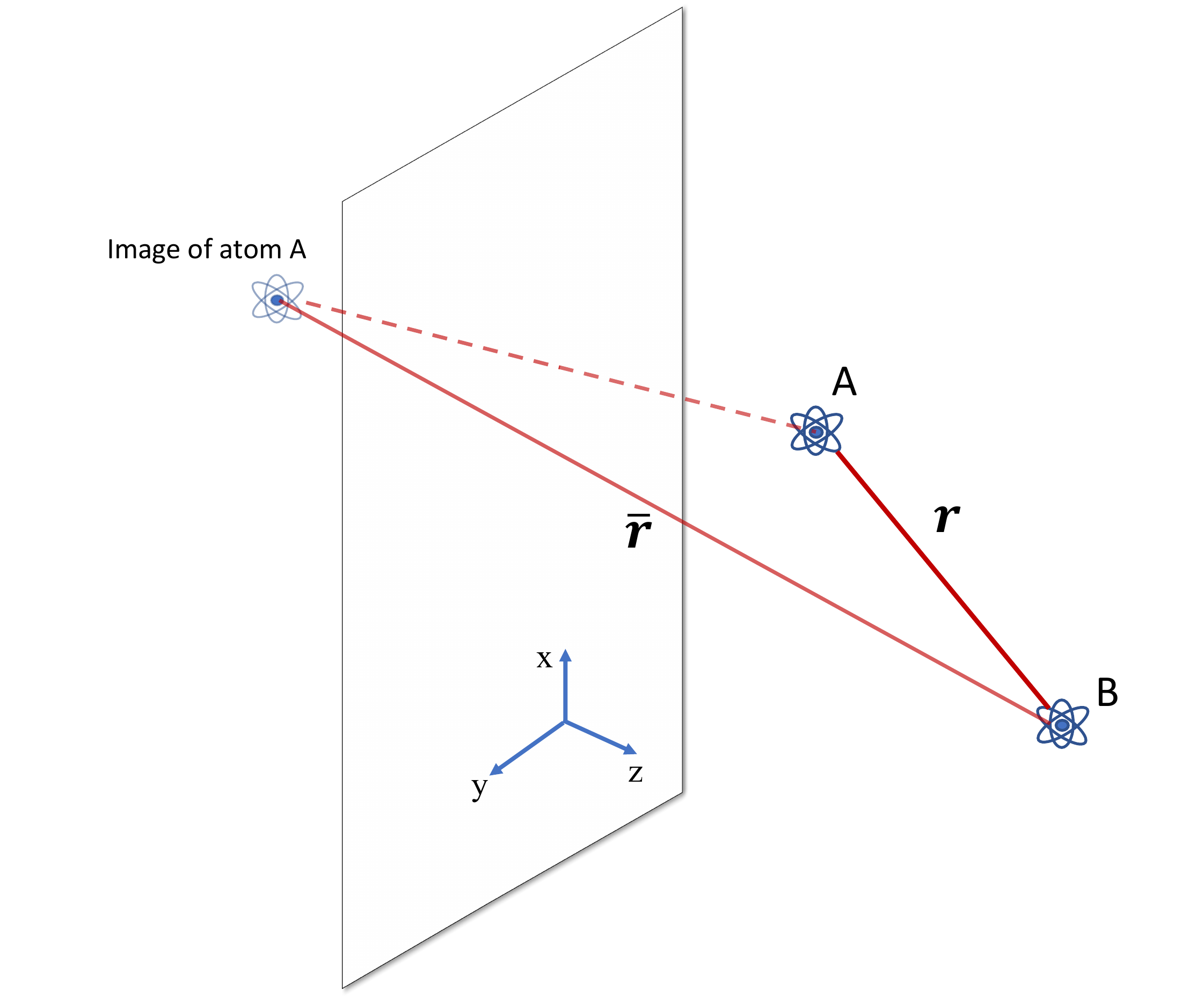}
\caption{Two atoms, A and B, nearby a reflecting plane boundary at $z=0$.}
\label{Fig:2}
\end{figure}

We consider two neutral atoms, A and B, in the vacuum space at zero temperature, near a perfectly conducting infinite plate located at $z=0$; $\br_A$ and $\br_B$ are respectively the positions of atoms A and B. We have already seen in Section \ref{sec:vacuum} that the presence of the plate changes vacuum fluctuations; mathematically, this is due to the necessity of setting appropriate boundary conditions on the field operators at the plate surface. Due to the deep relation of dispersion interactions to field fluctuations, shown in Section \ref{sec:vf}, we must expect that also the van der Waals and Casimir--Polder force (and any other radiation-mediated interaction) will change due to the plate \cite{Milton13,Armata17}. A fourth-order perturbative calculation  has been done in Ref. \cite{PT82}, using the multipolar coupling Hamiltonian with the appropriate mode functions of the field in the presence of the infinite reflecting plate; in the far zone, the result is
\begin{eqnarray}
\label{CPplate}
\Delta E_\text{r} (r, \bar{r}) &=& -\frac {23\hbar c}{4\pi} \alpha_A \alpha_B \frac 1{r^7} -\frac {23\hbar c}{4\pi} \alpha_A \alpha_B \frac 1{\bar{r}^7}
+\frac {8\hbar c}\pi \frac {\alpha_A \alpha_B}{r^3 \bar{r}^3 (r+\bar{r})^5}
\nonumber \\
&\ & \times \left[ r^4 \sin^2 \theta +5 r^3 \bar{r} \sin^2 \theta + r^2 \bar{r}^2 (6+\sin^2 \theta +\sin^2 \bar{\theta})
+5r \bar{r}^3 \sin^2 \bar{\theta} +\bar{r}^4 \sin^2 \bar{\theta} \right] ,
\end{eqnarray}
where $\br = \br_B -\br_A$ is the distance between the atoms, $\bar{\br} = \br_B - \sigma \br_A$ is the distance of atom B from the image of atom A (at point $\sigma \br_A$) with respect to the plate, having defined the reflection matrix $\sigma = \text{diag} (1,1,-1)$; $\theta$ and $\bar{\theta}$ are, respectively, the angles of $\br$ and $\bar{\br}$ with respect to the normal to the plate. The~geometrical configuration is illustrated in Figure \ref{Fig:2}.
This well-known expression shows that the retarded Casimir--Polder potential between the two atoms consists of three terms: the potential between A and B, scaling as $r^{-7}$, as in absence of the plate; the potential between an atom and the reflected image of the other atom, scaling as $\bar{r}^{-7}$; a term involving both distances $r$ and $\bar{r}$.

We now show that the potential (\ref{CPplate}) can also be obtained in a simpler way through the same methods used in Section \ref{sec:vf} for atoms in the unbounded space,  based on dressed energy densities and vacuum field correleations, stressing new physical insights on the origin of this potential.

We use the effective Hamiltonian (\ref{EffHam}) with the electric field operator given by (\ref{elfieldbound}), using the appropriate mode functions $\bfm (\bk \lambda ,\br )$ for a single infinite perfectly reflecting wall \cite{Compagno1995,Milonni1994,PT82}. The outline of the calculation is the same as in the case of atoms in free space (see Section \ref{sec:endens}): we evaluate the renormalized dressed electric energy density due to one atom (A), and then its interaction energy with the other atom (B).  The dressed ground state of atom A, as if atom B were absent, is given by
\begin{equation}
\label{drgrstateplate}
\mid \gd_A \rangle = \mid g_A, \vac \rangle - \frac \pi V \sum_{\bk \lambda \bkp \lambda '} \alpha_A (k) \frac {\sqrt{kk'}}{k+k'} \bfm (\bk \lambda ,\br_A) \cdot \bfm (\bkp \lambda ',\br_A)
\mid g_A, 1_{\bk \lambda} 1_{\bkp \lambda} \rangle .
\end{equation}

We now evaluate the average value of the effective interaction Hamiltonian relative to atom B on this state, disregarding bare terms that do not depend on the atomic distances, which is \cite{SPR06}
\begin{equation}
\label{EnergyAB}
\Delta E_{AB} =-\frac 12 \sum_{\bk \lambda \bkp \lambda '} \alpha_B (k)
\langle \gd_A \mid \bE (\bk \lambda , \br_B) \cdot \bE (\bkp \lambda ', \br_B) \mid \gd_A \rangle .
\end{equation}

Using the expression (\ref{elfieldbound}) of the electric field operator with the appropriate mode functions $\bfm (\bk \lambda ,\br )$ relative to the reflecting plate, polarization sum in (\ref{EnergyAB}) yields
\begin{equation}
\label{polsumplate}
\sum_\lambda f_i (\bk \lambda ,\br_A) f_j (\bk \lambda ,\br_B) = \left( \delta_{ij} -\hat{k}_i \hat{k}_j \right) e^{i\bk \cdot (\br_A -\br_B)}
-\sigma_{i\ell} \left( \delta_{\ell j} -\hat{k}_\ell \hat{k}_j \right) e^{i\bk \cdot (\br_A - \sigma \br_B)} ,
\end{equation}
where $i$, $j$, $\ell$ are Cartesian components, and $\sigma$ is the reflection matrix defined above. A comparison with~(\ref{polarizsum}) immediately shows an extra term due to the reflecting plate. Explicit evaluation of (\ref{EnergyAB}) in the far zone ($\alpha_{A,B}(k) \simeq \alpha_{A,B}$), using (\ref{polsumplate}), after some algebra finally yields the correct result (\ref{CPplate}), originally obtained in Ref. \cite{PT82} from a fourth-order calculation. The present approach, besides stressing the role of (dressed) vacuum fluctuations and their modification due to the plate, has allowed to obtain the same result through a simpler first-order calculation with the effective Hamiltonian given in \mbox{Equation (\ref{EffHam}) \cite{SPR06}.}

In addition, the method based on vacuum spatial correlations of the electric field operator, outlined in Section \ref{vacfieldcorr}, can be extended to the case when boundaries are present. Following the approach used in Section \ref{vacfieldcorr}, we need to evaluate the following quantity
\begin{equation}
\label{EABplatecorr}
\Delta E_{AB}(r,\bar{r}) = \sum_{\bk \lambda} \langle \vac \mid E_i (\bk \lambda ,\br_A) E_j (\bk \lambda ,\br_B) \mid \vac \rangle V_{ij}(k,r,\bar{r}),
\end{equation}
where $\mid \vac \rangle$ is the vacuum state of the field when the plate is present, and $V_{ij}(k,r,\bar{r})$ is the potential tensor giving the interaction energy for oscillating (induced) dipoles. In the present case, however, the~expression of $V_{ij}(k,r,\bar{r})$ differs from (\ref{ReF}), valid for dipoles in the free space, because of the effect of the image dipoles. We thus take
\begin{equation}
\label{pottensplate}
V_{ij}(k,r,\bar{r}) = V_{ij}(k,r) - \sigma_{i\ell} V_{\ell j}(k,\bar{r}) ,
\end{equation}
where the second term takes into account the image dipole, reflected on the plate. Substitution of the appropriate spatial field correlation function (obtained using (\ref{polsumplate})) and (\ref{pottensplate})  into (\ref{EABplatecorr}), after algebraic calculations, finally yields  the correct expression (\ref{CPplate}) for the retarded dispersion for two atoms nearby the conducting plate \cite{SPR06}. This finding indicates that the method based on field correlations can be directly extended to the case when boundary conditions are present, provided the appropriate field correlation function is used and the images dipole(s) are included in the classical potential tensor. It~can be shown that this method is valid also for evaluating the dispersion interaction with boundary conditions and at a finite temperature \cite{PS07}. All this is relevant because these methods can provide very useful computational tools to evaluate dispersion interactions in complicated geometries, also allowing the possibility to change and manipulate radiation-mediated interactions between atoms or molecules through the environment. This is still more striking for the resonance interaction, where the exchange of real photons is also present \cite{Armata17,Incardone14,Notararigo18}.

In the system considered above, we have assumed that the boundary is a perfectly reflecting one. This is an idealised situation, of course, because any real dielectric or metal material is characterised by specific magnetodielectric properies. For example, a real metal can be described with the plasma or the Drude-Lorentz model \cite{Bordag09,Kittel04}. In the plasma model, the metal is transparent for frequencies above its plasma frequency $\omega_p$, and this can affect renormalized vacuum fluctuations and the dispersion interaction energy between atoms nearby the boundary. It is expected that in general the corrections to vacuum fluctuations can be relevant only for short distances from the boundary \cite{BP12}, and the same is also expected for atom--surface and atom--atom interactions, similarly to the Casimir effect for dielectrics, as obtained through the Lifshits formula \cite{Lifshits56,LP80}. Indeed, the van der Waals interaction between an atom and a metallic plasma surface has been calculated, showing relevant corrections due to the finite plasma frequency in the short distance regime \cite{BB76}. In addition, the dispersion interaction between a ground-state atom and a  dielectric surface, and between two atoms nearby a dielectric surface, have been evaluated in terms of the dielectric constant of the surface at imaginary frequencies, using the theory of electrical images \cite{McLachlan64}. In the case of the atom--atom interaction, a~strong dependence (suppression or enhancement) of the interaction on the geometry of the two atoms with respect to the surface has been found using the linear response theory \cite{CS96}; enhanced dispersion interaction between two atoms in a dielectric slab between two different dielectric media has also been found~\cite{MD05}.
The~effect of real boundaries on atom--surface and atom--atom dispersion interactions can be conveniently included through appropriate body-assisted quantization of the electromagnetic field, based on the Green's functions technique \cite{MLBJ95,GW96,DKW98,BuhmannI-2012,BuhmannII-2012,BBS12,Scheel15}. In this approach, medium-assisted bosonic operators for the field are introduced, which take into account all magnetodielectric (dispersive and dissipative) properties of the linear macroscopic boundaries through their frequency-dependent complex electric permittivity and magnetic permeability; then, expressions of dispersion interactions for atoms or molecules nearby a generic linear environment can be obtained \cite{DKW98,BW07,SB08,BuhmannI-2012}, as well as expressions of the intermolecular energy transfer \cite{DKW02}.

\section{Casimir--Polder and Resonance Interactions between Uniformly Accelerated Atoms}
\label{sec:acceleration}
The results reviewed in the previous sections have shown the deep relation between renormalized vacuum energy densities, and zero-point field fluctuations, with dispersion interactions, in particular in the retarded Casimir--Polder regime.
{All these results and effects have been obtained for atoms and boundaries at rest. New phenomena appear when they are set in motion. In the case of a uniformly accelerated motion, the well-known Fulling--Davies--Unruh effect predicts that the accelerated observer perceives vacuum fluctuations as a thermal bath with the Unruh temperature $T_U= \hbar a/2\pi ck_B$, where~$a$ is its proper acceleration and $k_B$ is the Boltzmann constant \cite{Unruh1976,Fulling73,Davies75,Crispino2008,Milonni1994}. This effect leads, for example, to a change of the Lamb shift for atoms in uniformly accelerated motion \cite{Audretsch95,Passante98}, as well as to spontaneous excitation of an accelerating atom due to its interaction with the Unruh thermal quanta~\cite{Audretsch94,ZY06,Calogeracos16}. For a boundary moving with a nonuniform acceleration (a single oscillating reflecting plate, or a metal cavity with an oscillating wall, for example), the dynamical Casimir effect occurs, consisting of the emission of pairs of real photons from the vacuum \cite{Moore1970,Dodonov2010}. Theoretical analysis of the dynamical Casimir effect involves quantization of the scalar or electromagnetic field with a moving boundary \cite{Moore1970,DK96,MMN98}. Analogous effects related to Casimir energies occur in the case of a metallic cavity with a movable wall, whose mechanical degrees of freedom are described quantum-mechanically, and thus subject to quantum fluctuations of its position \cite{Law94,Law95,BP13,AP15,Armata17}. Another relevant effect involving atoms in motion, extensively investigated in the literature, is the quantum friction of an atom moving at uniform speed parallel to a surface \cite{BuhmannII-2012,DMNM11,Barton2010a,Barton2010b,Intravaia2015}.
An important question that can be asked, and that we shall address in this section, is what is the effect of a non inertial motion in the vacuum space of two or more atoms on their dispersion interaction.
}
We should expect a change of the dispersion interaction between the atoms because of their accelerated motion; in fact, they perceive a different vacuum, equivalent to a thermal bath, and in view of the temperature dependence of the dispersion interactions, mentioned in Section \ref{sec:vdW-CP}, their interaction energy should change. However, the Unruh effect is a very tiny effect, and an acceleration of the order of $10^{20} \,  \mbox{m/s}^{2}$  is necessary to get a Unruh temperature around $1$ K. Spontaneous excitation  of a uniformly accelerated atom has been predicted \cite{Audretsch94}, as well as changes of the Lamb shift of atomic levels and of the atom--surface Casimir--Polder force \cite{Passante98,Rizzuto09,Rizzuto07}.
Although some experimental setups to detect the Unruh effect have been proposed \cite{Schutzhold2006}, this effect has not been observed yet (a related effect, the Hawking radiation has been recently observed in analogue systems, specifically in an acoustical black hole \cite{Steinhauer16}). The change of dispersion interactions due to a uniformly accelerated motion of the atoms, if observed, could provide a signature of the Unruh effect \cite{NP13}.

For the sake of simplicity, we consider two two-level atoms of frequency $\w0$, separated by a distance $z$, and interacting with the massless relativistic scalar field. For atoms at rest and at zero temperature,
the scalar Casimir--Polder interaction at zero temperature behaves as $z^{-2}$ in the near zone ($z \ll c/\w0$), and as $z^{-3}$ in the far zone ($z \gg c/\w0$). At a finite temperature $T$, after defining the thermal length $\rho^\text{therm}  =\hbar c/(2\pi k_BT)$ and assuming $z \gg c/\w0$, the interaction for $z \ll \rho^\text{therm}$ has the same behaviour as at zero temperature, with a subleading thermal correction proportional to $T^2/z$, both in the near and in the far zone. At very large distances, $z \gg \rho^\text{therm}$, the interaction energy is proportional to
$T/z^2$, thus scaling with the distance as in the near zone at zero temperature (this should also be compared to the case of atoms interacting with the electromagnetic field at a finite temperature, mentioned in Section \ref{sec:vdW-CP}). We note that the scaling with the distance $z$ is different compared with the cases considered in the previous sections because we are now considering the scalar rather than the electromagnetic field.

We describe the two identical atoms, A and B, as two-level systems of frequency $\w0$, using in the Dicke formalism \cite{Dicke54}. They interact with the relativistic massless scalar field $\phi (x)$, and move in the vacuum space with the same constant proper acceleration $a$ along the direction $x$, perpendicular to their distance (along $z$), so that their separation is constant.

The Hamiltonian of this system, in the comoving frame, which is the system in which the atoms are instantaneously at rest, is
\begin{equation}
\label{Hamiltacc}
H =\hbar \w0 S_z^A(\tau ) +\hbar \w0 S_z^B(\tau ) + \sum_\bk \hbar \wk a_\bk^\dagger a_\bk \frac {dt}{d\tau} +\lambda S_x^A(\tau ) \phi [ x_A(\tau )] +\lambda S_x^B(\tau ) \phi [ x_B(\tau )] ,
\end{equation}
where $S_i \, (i=x,y,z)$ are pseudospin operators, $\tau$ is the proper time, shared by both atoms in our hypothesis, $\lambda$ is a coupling constant, and $x_{A(B)}(\tau )$ are the atomic trajectories
\begin{eqnarray}
\label{trajectory}
&\ & t_A(\tau ) = \frac ca \sinh (\frac {a\tau}c) , \, \, \, x_A (\tau )=\frac {c^2}a \cosh (\frac {a\tau}c) ; \, \, \, y_A(\tau )=0; \, \, \, z_A(\tau )=z_A ,
\nonumber \\
&\ & t_B(\tau ) = \frac ca \sinh (\frac {a\tau}c) , \, \, \, x_B (\tau )=\frac {c^2}a \cosh (\frac {a\tau}c) ; \, \, \, y_B(\tau )=0; \, \, \, z_A(\tau )=z_A +z ,
\end{eqnarray}
with $\tau_A =\tau_B =\tau$ the common proper time of the atoms, and $z$ the (constant) distance between them~\cite{Rindler06,BirrellDavies82}. The massless scalar field operator is
\begin{equation}
\label{scalarfield}
\phi (\br ,t) = \sum_\bk \sqrt{\frac \hbar{2V\wk}} \left[ a_\bk e^{i(\bk \cdot \br -\wk t)}+ a_\bk^\dagger e^{-i(\bk \cdot \br -\wk t)} \right] ,
\end{equation}
with the dispersion relation $\wk = c\mid \bk \mid$.
The Casimir--Polder energy is the distance-dependent energy shift due to the atoms-field interaction, and it is a fourth-order effect in the coupling constant $\lambda$.
The~calculation of the relevant part of the energy shift can be done using a method, originally introduced by Dupont-Roc et al. to separate vacuum fluctuations and radiation reaction contributions to second-order radiative shifts \cite{Dalibard1982,Dalibard1984}, even in accelerated frames \cite{Zhou2012,Menezes2016}, and recently generalized to fourth-order processes \cite{MNP14,NMP18}. We find that a new distance scale, related to the acceleration, $z_a = c^2/a$, appears. For $z \ll z_a$, the dispersion interaction for atoms with the same uniform acceleration $a$ is the same as that for atoms at rest in a thermal bath with the Unruh temperature $T_U= \hbar a/2\pi ck_B$. In~this regime, our extended system of two atoms exhibits the Unruh equivalence between acceleration and temperature, similarly to a point-like detector. However, for larger distances, $z \gg z_a$, we have
\begin{equation}
\label{CPacc}
\Delta E^\text{accel} \simeq -\frac {\lambda^4}{512 \pi^4 \hbar c \w0^2 a}\frac 1{z^4}.
\end{equation}

In this regime, the interaction energy decays with the distance faster than in the near and far zone, showing lack of the equivalence between acceleration and temperature \cite{MNP14}. In addition, the dispersion interaction between accelerating atoms has qualitative features, specifically its distance dependence, different from those of inertial atoms; measure of the dispersion force between atoms subjected to a uniform acceleration could thus give an indirect signature of the Unruh effect.

The non-thermal behaviour of acceleration at large distances is obtained also for the resonant interaction energy between two identical accelerating atoms, one excited and the other in the ground state, prepared in an entangled symmetric or antisymmetric Bell-type state
\begin{equation}
\label{corrstate}
\mid \phi_\pm \rangle = \frac 1{\sqrt{2}} \left( \mid e_A g_B , \vac \rangle \pm \mid g_A e_B , \vac \rangle \right) ,
\end{equation}
where $e_{A(B)}$ and $g_{A(B)}$ respectively represent the excited and the ground state of the atoms. In this case, a {\it resonance} interaction energy between the two correlated atom exists, due to the exchange of one real or virtual photon between them.

We consider two identical two-level atoms, A and B, with frequency $\w0=ck_0$, and interacting with the electromagnetic field through the multipolar coupling Hamiltonian.
$\bmu$ is the dipole moment operator. In the unbounded space, at zero temperature, the resonance interaction energy for atoms at rest, obtained by evaluating the second-order energy shift due to the interaction Hamiltonian (\ref{HamiltonianCP}), is~given by
\begin{equation}
\label{resint}
\Delta E ^\text{res}_\pm = \pm \mu_{Ai}^{eg} \mu_{Bj}^{ge} \left[ \left( \delta_{ij}-3\hat{r}_i \hat{r}_j \right) (\cos (k_0r) + k_0r \sin (k_0r))- \left( \delta_{ij}-\hat{r}_i \hat{r}_j \right)
k_0^2 r^2 \cos (k_0r) \right] \frac 1{r^3} ,
\end{equation}
where the $+$ or $-$ sign respectively refers to the symmetric or antisymmetric state in (\ref{corrstate}), $r$ is the distance between the atoms, and $\mu_{A(B)i}^{eg}$ is the matrix element of $(\mu_{A(B)})_i$ between the excited and the ground state, assumed real. In the near zone ($k_0r \ll 1$), the interaction energy (\ref{resint}) scales as $r^{-3}$, while~in the far zone ($k_0r \gg 1$) it scales as $r^{-1}$ with space oscillations \cite{Craig1998,Incardone14}.

We now assume that the two atoms are moving with a uniform proper acceleration along $x$, and~that their distance is along $z$, so that it is constant in time; their trajectory is given by Equations (\ref{trajectory}). The Hamiltonian, similarly to Equation (\ref{Hamiltacc}) of the scalar field case, within dipole approximation and in the comoving frame, is
\begin{equation}
\label{Ham-accel-em}
H = H_A + H_B + \sum_{\bk \lambda} \hbar \wk \akld \akl \frac {dt}{d\tau} - \bmu_A(\tau ) \cdot \bE (x_A(\tau )) - \bmu_B(\tau ) \cdot \bE (x_B(\tau )) ,
\end{equation}
where $H_{A(B)}$ is the Hamiltonian of atom A (B), and $\bmu_{A(B)}$ is their dipole moment operator.

After lengthy algebraic calculations, for $z \gg z_a = c^2/a$, the following expression for the resonant interaction energy between the two accelerating atoms, in the comoving frame, is obtained
\begin{eqnarray}
\label{resintaccel}
\Delta E ^\text{accel}_\pm &\simeq& \pm \mu_{A\ell}^{eg} \mu_{Bm}^{ge} \frac 1{z^3} \Bigg\{ (\delta_{\ell m} -q_\ell q_m -2 n_\ell n_m)
\Big[ \frac {2\w0 z}c \sin \left( \frac {2\w0 c}a \log \left( \frac {az}{c^2} \right) \right)
\nonumber \\
&\ &  - \frac {\w0^2z^2}{c^2} \left( \frac {2c^2}{za} \right) \cos \left( \frac {2\w0 c}a \log \left( \frac {az}{c^2} \right) \right) \Big]
+ q_\ell q_m \left( \frac {8c^2}{az} \right)
\cos \left( \frac {2\w0 c}a \log \left( \frac {az}{c^2} \right) \right) \Bigg\} ,
\end{eqnarray}
where $\ell ,m$ are Cartesian components, ${\bf n} = (0,0,1)$ is a unit vector along $z$ (the direction of the distance between the atoms), ${\bf q} = (1,0,0)$ is a unit vector along $x$ (the direction of the acceleration), and $\pm$ refers to the symmetric or antisymmetric superposition in (\ref{corrstate})  \cite{Rizzuto2016}. Comparison of (\ref{resintaccel}) with (\ref{resint}) shows that the acceleration of the atoms can significantly change the distance-dependence of the resonance interaction, with respect to inertial atoms. For atoms at rest, Equation (\ref{resint}) gives a far-zone dependence as $z^{-1}$, while for accelerating atoms Equation (\ref{resintaccel}) asymptotically gives a more rapid decrease of the interaction with the distance: $z^{-2}$ if the two dipoles are along $z$ or $y$, and $z^{-4}$ for dipole oriented along the direction of the acceleration, $x$. This change in the dependence of the interaction energy from the interatomic distance is a signature of the noninertial motion of the atoms, and ultimately related to the Unruh effect (even if, in this case, the change of the interaction energy is not equivalent to that due to a thermal field). Moreover, Equation (\ref{resintaccel}) shows striking features of the accelerated-atoms case with respect to the orientation of the atomic dipole moments. For example, if the two dipoles are orthogonal to each other, with one along $z$ and the other in the plane $(x,y)$, the interaction energy vanishes for inertial atoms, as immediately follows from (\ref{resint}); on the other hand, Equation  (\ref{resintaccel}) shows that it is different from zero when $a \neq 0$. In such a case, the resonance interaction is a unique signature of the accelerated motion of the atoms  \cite{Rizzuto2016}. Similar results are also found in the coaccelerated frame \cite{Zhou2016}, and for atoms nearby a reflecting boundary \cite{Zhou2018}.  Finally, we wish to mention that very recently the Casimir--Polder and resonance interactions between atoms has been also investigated in the case of a curved-spacetime background \cite{Menezes17,Zhou2017,Zhou2018a}.

In conclusion, the results outlined in this section give a strong indication that the radiation-mediated interactions, specifically dispersion and resonance interactions, between uniformly accelerated atoms could provide a promising setup to detect the effect of a noninertial motion on radiative processes, possibly allowing an indirect detection of the Unruh effect through the measurement of dispersion or resonance interaction between accelerating atoms \cite{NP13,MNP14,Rizzuto2016,Zhou2018}.

\section{Conclusions}
\label{sec:summary}
In this review, we have discussed recent developments on radiation-mediated interactions between atoms or molecules, in the framework of nonrelativistic quantum electrodynamics. Major emphasis has been on stressing the role of zero-point field fluctuations and (dressed) vacuum field energy densities and spatial correlations. This has allowed us to give a transparent physical interpretation of dispersion (van der Waals and Casimir--Polder) and resonance interactions, as well as useful computational tools for their evaluation, even in more complicated situations, for example in the presence of boundaries or for uniformly accelerating atoms. We have shown that dispersion interactions can be seen as a direct consequence of the existence of vacuum fluctuations, with features directly related to their physical properties. We have also discussed how dispersion and resonance interactions could provide an experimental setup to get an indirect evidence of the Unruh effect, also testing the Unruh acceleration-temperature equivalence.

A fundamental final question arises: are van der Waals and Casimir--Polder dispersion interactions a definitive proof of the real existence of vacuum fluctuations and zero-point energy? In the author's opinion, the answer is yes and no. As clearly shown by all physical systems and situations discussed in this review, the results here reviewed are certainly fully consistent with the existence of vacuum fluctuations. Assuming the existence of vacuum fluctuations of the electromagnetic field, as predicted by nonrelativistic quantum electrodynamics, we can derive these observable phenomena, using clear and transparent physical models. However, the same results can be also obtained in a different way, that is, from source fields, without invoking vacuum fluctuations \cite{Milonni82}. A similar consideration applies to the Lamb shift, for example \cite{Milonni1994}.
This approach is based on the solution of the Heisenberg equations of motion for the field operators, which contains a free (vacuum) term and a source term, the latter depending on the presence of matter, while the former is related to the vacuum field. Using a specific ordering, specifically the normal ordering, between field and atomic operators (while the atomic operators commute with the full field operators, in general they do not commute with the single parts of field operators related to vacuum and source terms \cite{Senitzky73,MAS73}); it is possible to show that only the radiation reaction term contributes to dispersion interactions when this ordering of operators is chosen \cite{Milonni82,Milonni76}.
Thus, it seems that a deep dichotomy between vacuum fluctuations and source fields exists in nature. As pointed out and stressed by P.W. Milonni, the two physical models and interpretations of several radiative processes, in terms of vacuum fluctuations or source fields (radiation reaction), should be considered as the {\it two sides of a coin} \cite{Milonni88}. In conclusion, we just wish to mention that, probably, a~definitive and unambiguous confirmation of the existence of the zero-point energy (density) can be obtained only from its gravitational effects, it being a component of the energy-momentum tensor, and thus a source term for the gravitational field \cite{Adler95,Cree2018,Sola2013}. The essential point is that Casimir and Casimir--Polder forces are always related to {\em differences} of vacuum energies for different configurations of the system; necessarily, this involves the presence of matter and thus of both vacuum and source fields. On the contrary, gravitational effects are related to the absolute value of the vacuum energy density \cite{Carroll2014}. In any case, as we have pointed out above, dispersion interactions between atoms, also~in the presence of macroscopic bodies or for accelerated systems, are observable physical effects fully consistent with the real existence of the zero-point energy of the quantum electromagnetic theory, even if they cannot be considered as a definitive proof of the existence of the vacuum energy.

\vspace{10pt}


\noindent
{\bf Funding:} This research received no external funding.

\vspace{6pt}

\noindent
{\bf Acknowledgments:}
The content of this review, and many ideas therein, owe much to the long-standing past collaboration of the author with Franco Persico, Edwin A. Power and Thiru Thirunamachandran, to whose memory this
paper is dedicated. The author also wishes to thank Lucia Rizzuto for a careful and critical reading of the manuscript. The~author gratefully acknowledges financial support from the Julian Schwinger Foundation.

\vspace{6pt}

\noindent
{\bf Conflicts of interest:} The author declares no conflict of interest.

\end{document}